\documentclass[twocolumn]{autart}
\usepackage{mathtools}
\usepackage{amssymb}
\usepackage{balance}
\usepackage{enumitem}
\usepackage{mathdots}
\usepackage{xfrac}
\usepackage{bm}
\usepackage{array}
\usepackage{algorithm}
\usepackage{algorithmicx}
\usepackage{algpseudocode}

\newtheorem{assumption}{Assumption}
\usepackage{color}
\definecolor{mygreen}{RGB}{70, 156, 95}
\usepackage{etoolbox}

\makeatletter
\def\@xnamedef#1{\expandafter\protected@xdef\csname #1\endcsname}
\def\no@harm{} 
\def\ead@au#1{\protected@edef\@ead@au{#1}}
\patchcmd\runningauthor@fmt{\global\edef}{\protected@xdef}{}{}
\patchcmd\runningauthor@fmt{\global\edef}{\protected@xdef}{}{}
\patchcmd\author@fmt{\edef}{\protected@edef}{}{}
\patchcmd\add@xtok{\xdef}{\protected@xdef}{}{}
\makeatother
                                           
\usepackage{color}
\usepackage{eso-pic}

\AddToShipoutPictureBG*{%
	\AtPageUpperLeft{%
		\setlength\unitlength{1in}%
		\makebox(8,-1.75)[c]{
			\begin{tabular}{c c}
				Rodrigo A. Gonz\'alez \emph{et al.},
				Consistency analysis of refined instrumental variable methods for continuous-time\\
				system identification in closed-loop. To appear in
				{\em Automatica},
				2024,
				uploaded to ArXiv on April 13th, 2024 \\
\end{tabular}}}}

\begin{document}
\begin{frontmatter}
\title{Consistency analysis of refined instrumental variable methods for continuous-time system identification in closed-loop\thanksref{footnoteinfo}}

\thanks[footnoteinfo]{This paper was not presented at any IFAC 
meeting. Corresponding author: R.~A.~González.}

\author[TUE]{Rodrigo A. Gonz\'alez}\ead{r.a.gonzalez@tue.nl},    
\author[UON]{Siqi Pan}\ead{siqi.pan@uon.edu.au},               
\author[KTH]{Cristian R. Rojas}\ead{crro@kth.se},    
\author[UON]{James S. Welsh}\ead{james.welsh@newcastle.edu.au}  

\address[TUE]{Control Systems Technology Section, Department of Mechanical Engineering, Eindhoven University of Technology, Eindhoven, The Netherlands.} 
\address[UON]{School of Engineering, University of Newcastle, Callaghan, 2308 NSW, Australia} 
\address[KTH]{Division of Decision and Control Systems, KTH Royal Institute of Technology, 10044 Stockholm, Sweden}  


\begin{abstract}
Refined instrumental variable methods have been broadly used for identification of continuous-time systems in both open and closed-loop settings. However, the theoretical properties of these methods are still yet to be fully understood when operating in closed-loop. In this paper, we address the consistency of the simplified refined instrumental variable method for continuous-time systems (SRIVC) and its closed-loop variant CLSRIVC when they are applied on data that is generated from a feedback loop. In particular, we consider feedback loops consisting of continuous-time controllers, as well as the discrete-time control case. This paper proves that the SRIVC and CLSRIVC estimators are not generically consistent when there is a continuous-time controller in the loop, and that generic consistency can be achieved when the controller is implemented in discrete-time. Numerical simulations are presented to support the theoretical results.
\end{abstract}
\end{frontmatter}

\section{Introduction}
System identification deals with obtaining mathematical models of systems given measured data. Much has been written about this topic \cite{ljung1998system,soderstrom1988system}, and special focus has been given to the nature of the models in time (discrete-time or continuous-time \cite{garnier2008book}), as well as to the system setting (open-loop or closed-loop). Even though the system identification community concentrates most of its energy in modeling in discrete-time, there are reasons why continuous-time systems are more convenient to estimate: advantages can be found in the robustness and parsimony of the models, physical insight of the estimated parameters, and lack of dependency of the models on the sampling period, among other attributes \cite{garnier2014advantages}.

Instrumental variables have been extensively used for system identification in closed-loop settings. Methods based on this approach have been developed and discussed for discrete-time models in closed-loop \cite{gilson2005instrumental,gilson2011optimal}, and an extension for discrete-time Hammerstein models can be found in \cite{laurain2009refined}. For continuous-time models, one option is to disregard the feedback and perform direct system identification using the refined instrumental variable method for continuous-time systems (RIVC), or its simplified embodiment, the SRIVC method~\cite{young1980refined}. Although the SRIVC estimator has been analyzed extensively in open-loop \cite{pan2020consistency,pan2020efficiency,gonzalez2020consistent}, several questions still remain unsolved regarding its statistical properties in closed-loop settings. On the other hand, various algorithms have been proposed for incorporating the controller into the estimation process. A bias-eliminated least-squares method was derived in \cite{garnier2000bias}, while \cite{gilson2003continuous} explored instrumental variable techniques. These were later revisited in \cite{young2009simple} and \cite{young2008refined}, where the SRIVC and RIVC methods are implemented using simulation data for obtaining a model estimate in closed-loop. A three-stage estimator was detailed in~\cite{young2009three}, and another two-stage algorithm was introduced in \cite{li2015closed}. An extensive review of instrumental variable methods for closed-loop continuous-time system identification can be found in \cite{gilson2008instrumental}, in which the authors introduce a closed-loop variant of SRIVC called CLSRIVC (closed-loop SRIVC) that requires the knowledge of the continuous-time controller. This estimator is argued to be asymptotically unbiased for white and colored output noise \cite[pp. 154-156]{garnier2008book}, although no proof of this statement is given.

There are two setups that are commonly addressed in the contributions cited above, which differ in the nature of the controller (continuous-time or discrete-time). This paper analyzes the consistency of the SRIVC and CLSRIVC methods when they are applied with continuous-time and discrete-time controllers in the feedback loop, and clarifies the conditions that are sufficient for obtaining consistent estimates of the continuous-time system under study. In summary, the main contributions of this paper are:
\begin{itemize}
	\item
	we prove that the SRIVC and CLSRIVC estimators are not consistent (and in particular, not asymptotically unbiased) when the data is generated from a fully continuous-time feedback loop and only sampled data are available as measurements. The lack of consistency of the CLSRIVC estimator in this scenario has been overlooked in the literature (see, e.g., \cite[p. 154]{gilson2008instrumental}) where it is claimed that CLSRIVC is asymptotically unbiased;
	\item
	we illustrate that the lack of asymptotic unbiasedness of the methods can be mitigated via oversampling techniques;
	\item
	we show that, under some technical assumptions, the SRIVC estimator is generically consistent when the controller in the loop is in discrete-time. However, consistency is lost if there is a direct feedthrough in the loop transfer or if the noise is colored. In such cases, analogously to some discrete-time settings \cite{welsh2002finite}, the discrete-time equivalent of the model estimate is shown to be biased towards the negative inverse of the controller; and
	\item
	we derive a variant of the CLSRIVC estimator that is proven to be generically consistent when the controller works in discrete-time and is known.
\end{itemize}

This paper is organized as follows. Section \ref{closedloopsetups} introduces the two feedback settings we study, and the SRIVC and CLSRIVC estimators are presented in Section \ref{sec:SRIVC}. Sections \ref{setup1} and \ref{setup2} contain the asymptotic properties of the SRIVC and CLSRIVC methods for the feedback loops of Settings 1 and 2 respectively, as shown in Fig. \ref{fig1}. Simulations that support our analysis can be found in Section \ref{simulations}, while conclusions are drawn in Section \ref{conclusions}. 

\section{Closed-loop settings}

\label{closedloopsetups}
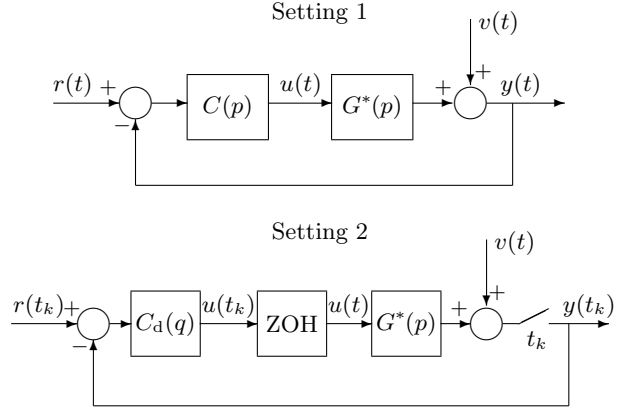
\begin{figure}
	\setlength{\unitlength}{0.105in} 
	\centering 
	\begin{picture}(30,20.5) 
		\put(2.1,15.1){\vector(1,0){3.3}}
		\put(2.1,15.7){\footnotesize{$r(t)$}}
		\put(13,19.3){\small{Setting 1}}
		\put(4.4,15.6){\scriptsize{+}}
		\put(5.1,13.7){\scriptsize{$-$}}
		\put(6.2,11){\vector(0,1){3.3}}
		\put(6.2,11){\line(1,0){18.8}}
		\put(6.2,15.1){\circle{1.6}}
		\put(7,15.1){\vector(1,0){1.8}}
		\put(8.8,13.3){\framebox(4,3.2){\small{$C(p)$}}}
		\put(12.8,15.1){\vector(1,0){3.2}}
		\put(13.4,15.7){\footnotesize{$u(t)$}}
		\put(16,13.3){\framebox(4,3.2){\footnotesize{$G^*\hspace{-0.02cm}(p)$}}}
		\put(20,15.1){\vector(1,0){2}}
		\put(21.1,15.6){\scriptsize{+}}
		\put(23.3,18.8){\footnotesize{$v(t)$}}
		\put(22.9,15.1){\circle{1.6}}
		\put(22.9,19.3){\vector(0,-1){3.3}}
		\put(23.1,16.4) {\scriptsize{+}}
		\put(25,11){\line(0,1){4.1}}
		\put(23.7,15.1){\vector(1,0){3.9}}
		\put(24.4,15.7){\footnotesize{$y(t)$}}	
		
		\put(0,4.1){\vector(1,0){3.3}}
		\put(0.1,4.7){\footnotesize{$r(t_k)$}}
		\put(2.6,4.6){\scriptsize{+}}
		\put(13,8.3){\small{Setting 2}}
		\put(3,2.7){\scriptsize{$-$}}
		\put(4.1,0){\vector(0,1){3.3}}
		\put(4.1,0){\line(1,0){23.7}}
		\put(4.1,4.1){\circle{1.6}}
		\put(4.9,4.1){\vector(1,0){1.1}}
		\put(6,2.5){\framebox(3.4,3.2){\small{$C_\textnormal{d}(q)$}}}
		\put(9.4,4.1){\vector(1,0){2.9}}
		\put(9.5,4.7){\footnotesize{$u(t_k)$}}
		\put(12.3,2.5){\framebox(3.4,3.2){\footnotesize{ZOH}}}
		\put(15.7,4.1){\vector(1,0){2.3}}
		\put(15.8,4.7){\footnotesize{$u(t)$}}
		\put(18,2.5){\framebox(3.4,3.2){\footnotesize{$G^*\hspace{-0.02cm}(p)$}}}
		\put(21.4,4.1){\vector(1,0){1.5}}
		\put(21.9,4.6){\scriptsize{+}}
		\put(24.1,7.8){\footnotesize{$v(t)$}}
		\put(23.7,4.1){\circle{1.6}}
		\put(23.7,8.3){\vector(0,-1){3.3}}
		\put(23.8,5.4) {\scriptsize{+}}
		\put(24.5,4.1){\line(1,0){0.8}}
		\put(25.3,4.1){\line(2,1){1.5}}
		\put(25.7,3){\footnotesize{$t_k$}}	
		\put(27.8,0){\line(0,1){4.1}}
		\put(26.8,4.1){\vector(1,0){3}}
		\put(27.5,4.7){\footnotesize{$y(t_k)$}}	
	\end{picture}
	\vspace{0.1cm}
	\caption{Block diagrams for the closed-loop Settings 1 and 2.}
	\label{fig1}
\end{figure}

Consider a linear, time-invariant, causal, single-input single-output, continuous-time system $G^*(p)=B^*(p)/A^*(p)$, where
\begin{align}
	B^*(p) &= b_{m^*}^*p^{m^*}+\cdots + b_{1}^*p + b_0^*, \notag \\
	A^*(p) &= a_{n^*}^*p^{n^*}+\cdots + a_{1}^*p + 1, \notag
\end{align}
with the polynomial degrees satisfying $n^*\geq m^*$, and $p$ being the Heaviside operator. The zero-order hold (ZOH) equivalent discrete-time system of $G^*(p)$ is denoted by $G_\textnormal{d}^*(q)$, where $q$ is the forward-shift operator. Given sampled data of the input and output signals, the main goal is to estimate the parameters of the transfer function $G^*(p)$ that are described by
$\bm{\theta}^* = [a_1^*, \hspace{0.1cm} a_2^*, \hspace{0.1cm} \dots, \hspace{0.1cm} a_{n^*}^*, \hspace{0.1cm} b_0^*, \hspace{0.1cm}b_1^*, \hspace{0.1cm}\dots, \hspace{0.1cm} b_{m^*}^*]^\top$. In this paper we study the generic consistency of the SRIVC and CLSRIVC estimators for the two closed-loop settings presented in Figure \ref{fig1}. In both scenarios we assume that the reference signal is reconstructed from a ZOH device (i.e., it is constant between samples).\footnote{This assumption is stated for simplicity of notation only; our results can extend to the first-order-hold and continuous-time multisine cases with some minor modifications by following the reasoning in \cite[Cor. 2]{pan2020consistency} and \cite[Thm. 4]{gonzalez2020consistent}.} A detailed description of each setting is given below.

\begin{itemize}
	\item
	\textit{Setting 1:} 
	The first setting assumes that the controller is a continuous-time transfer function $C(p)=F(p)/L(p)$ and that the system output is corrupted by a measurement noise $v(t)$. Since we do not have access to the noise behavior between the sampling time instants of the output, the most convenient option is to model the noise as being constant between samples and that the noise samples $v(t_k)$ that perturb the output are a realization of some stochastic process. This assumption is used in \cite{gilson2008instrumental}, as well as in  \cite{li2015closed,victor2017closed,gonzalez2021unstable}. Other contributions have not used this assumption \cite{garnier2000bias,gilson2003continuous}, although limited analysis has been done in such context. In this work, the disturbance signal is assumed to be constant between the samples of the output. We note that our theoretical results could be extended for the continuous-time white noise case, but extra care must be taken to avoid the theoretical problems associated with continuous-time white noise~\cite{aastrom1970introduction}.
	
	Under Setting 1, we can express $u(t_k)$ and $y(t_k)$ as a function of the signals $r(t_k)$ and $v(t_k)$ as follows:
	\begin{subequations}
		\label{setup1eq}
		\begin{align}
			\label{utk1}
			u(t_k) &= S_{uo}^*(p) r(t_k) - S_{uo}^*(p)v(t_k)  \\
			\label{ytk1}
			y(t_k) &= T_o^*(p)r(t_k) + S_o^*(p)v(t_k),
		\end{align}
	\end{subequations}
	where we have used the standard notation of sensitivity functions \cite{goodwin2001control}
	\begin{align}
		T_o^*(p)\hspace{-0.05cm}&:= \hspace{-0.04cm}\frac{G^*(p)C(p)}{1\hspace{-0.03cm}+\hspace{-0.03cm}G^*(p)C(p)}, \hspace{0.1cm} S_o^*(p)\hspace{-0.05cm}:=\hspace{-0.04cm}\frac{1}{1\hspace{-0.03cm}+\hspace{-0.03cm}G^*(p)C(p)}, \notag \\
		\label{sensitivity}
		S_{uo}^*(p)\hspace{-0.05cm}&:= \hspace{-0.04cm}\frac{C(p)}{1\hspace{-0.03cm}+\hspace{-0.03cm}G^*(p)C(p)}.
	\end{align}
	\begin{rem}\hspace{-0.2cm}\textbf{.}
		\label{remarkmixed}
		In \eqref{utk1} and \eqref{ytk1} we have introduced a mixed notation of continuous-time filters with discrete-time signals. If $G(p)$ is a continuous-time filter, and $x(t_k)$ is a sampled signal, then $G(p)x(t_k)$ implies that the signal $x(t_k)$ is interpolated using either a zero or first-order hold (ZOH or FOH), and the resulting output of the filter is sampled at $t_k$. On the other hand, the expression $\{G(p)x(t)\}_{t=t_k}$ (or $[G(p)x(t)]_{t=t_k}$ in the vector-valued case) means that the continuous-time signal $x(t)$ is filtered through $G(p)$ and later sampled at $t=t_k$. For example, if the zero-order hold equivalents of $G_1(p)$, $G_2(p)$ and $G_1(p)G_2(p)$ are $H_1(q)$, $H_2(q)$ and $H_{12}(q)$ respectively, and $r(t)$ is constant between samples, then $G_1(p)\{G_2(p)r(t)\}_{t=t_k}=[H_1(q)H_2(q)]r(t_k)$, while $[G_2(p)G_1(p)]r(t_k)=H_{12}(q)r(t_k)$. Note that in general we have $H_1(q)H_2(q)\neq H_{12}(q)$.
	\end{rem}
	
	\item
	\textit{Setting 2:} The second setting is hybrid in nature, as it considers a discrete-time controller $C_\textnormal{d}(q)=F_\textnormal{d}(q)/L_\textnormal{d}(q)$ acting upon a continuous-time system. The controller output $u(t_k)$ is a discrete-time signal that is reconstructed through a ZOH device before acting on $G^*(p)$, and the noisy system output is sampled before being fed back to the controller. This framework is used in \cite{young2009simple,young2008refined,young2009three}.
	
	In this setting the input and output satisfy
		\begin{align}
			u(t_k) &= S_{uo}^*(q) r(t_k) - S_{uo}^*(q)v(t_k) \notag \\
			y(t_k) &= T_o^*(q)r(t_k) + S_o^*(q)v(t_k), \notag 
		\end{align}
	where the sensitivity functions are in the same form as those in \eqref{sensitivity}, but with $G^*(p)$ and $C(p)$ being replaced by $G_\textnormal{d}^*(q)$ and $C_\textnormal{d}(q)$ respectively.
\end{itemize}
Both settings assume that the samples $\{u(t_k),y(t_k)\}_{k=1}^N$ (and possibly $\{r(t_k)\}_{k=1}^N$) are available as data for identification, and we define $n_f^*$ and $n_l^*$ as the number of zeros and poles respectively of $C(p)$ or $C_\textnormal{d}(q)$. The general assumptions that we require for our analysis are:
\begin{assumption}
	\label{assumption21}
	The true system $B^*(p)/A^*(p)$ is proper $(n^*\geq m^*)$ and asymptotically stable with $A^*(p)$ and $B^*(p)$ being coprime. The closed-loop system is also asymptotically stable, i.e., the zeros of $1+G^*(p)C(p)$ are in the open left half plane for Setting 1, and the zeros of $1+G_\textnormal{d}^*(q)C_\textnormal{d}(q)$ are in the open unit disk for Setting 2.
\end{assumption}
\begin{assumption}
	\label{assumption22}
	The external reference $r(t_k)$ and disturbance $v(t_l)$ are stationary and mutually independent for all integers $k$ and $l$.
\end{assumption}
\begin{assumption}
	\label{assumption23}
	The degrees $n$ and $m$ of the polynomials in the model satisfy $\min(n-n^*,m-m^*)=0$. For Setting 1 we also require $n-n^*\leq n^*-m^*+n_l^*-n_f^*$, whereas for Setting 2 we also assume that $n-n^*\leq n^*-m^*$.
\end{assumption}
The SRIVC and CLSRIVC estimators cannot directly handle the identification of unstable systems due to the prefiltering procedure becoming unstable \cite{young2012recursive,gonzalez2021unstable}. Thus, the stability of the system in open loop is needed for the filtering steps in both algorithms. Although it has been shown in \cite{gonzalez2021unstable} that this condition is no longer required if a more involved filtering technique is implemented, we do not pursue an analysis of such refinements here. We also note that the assumption on the model order is similar to Assumption 5 of \cite{pan2020consistency} for the consistency analysis of the SRIVC estimator in open loop, and it is used in the same manner in our proofs.

\section{The SRIVC and CLSRIVC estimators}
\label{sec:SRIVC}
The SRIVC estimator \cite{young1980refined} is an iterative instrumental variable method that generates parameter-dependent filters that are applied to sampled input and output data at each iteration. Its closed-loop version is called the CLSRIVC estimator \cite{garnier2008book}. The iterations of the SRIVC and CLSRIVC procedures require updating a filtered regressor vector $\bm{\varphi}_f(t_k,\bm{\theta}_j)$, a filtered instrument vector $\hat{\bm{\varphi}}_f(t_k,\bm{\theta}_j)$, and a filtered output $y_f(t_k,\bm{\theta}_j)$. Once these filtered signals are computed, the model estimate of the next iteration is given by the instrumental variable step
\begin{equation}
	\label{iterations}
	\bm{\theta}_{j\hspace{-0.02cm}+\hspace{-0.02cm}1}\hspace{-0.11cm}=\hspace{-0.11cm} \left[\sum_{k=1}^N \hspace{-0.09cm}\hat{\bm{\varphi}}_{\hspace{-0.02cm}f}\hspace{-0.02cm}(\hspace{-0.01cm}t_k\hspace{-0.02cm},\hspace{-0.03cm}\bm{\theta}_j\hspace{-0.02cm}) \bm{\varphi}_{\hspace{-0.02cm}f}^{\hspace{-0.02cm}\top}\hspace{-0.06cm}(\hspace{-0.01cm}t_k\hspace{-0.02cm},\hspace{-0.03cm}\bm{\theta}_j\hspace{-0.02cm})\hspace{-0.03cm}\right]^{\hspace{-0.08cm}-\hspace{-0.02cm}1}\hspace{-0.13cm}\left[\sum_{k=1}^N  \hspace{-0.09cm}\hat{\bm{\varphi}}_{\hspace{-0.02cm}f}\hspace{-0.02cm}(\hspace{-0.01cm}t_k,\hspace{-0.03cm}\bm{\theta}_j\hspace{-0.02cm}) y_f\hspace{-0.02cm}(\hspace{-0.01cm}t_k,\hspace{-0.03cm}\bm{\theta}_j\hspace{-0.02cm})\hspace{-0.02cm}\right]\hspace{-0.1cm}.
\end{equation} 
These iterations are performed until a fixed maximum number of iterations is reached or until the relative error between the previous and current
estimates is smaller than a preset tolerance factor $\epsilon$. To avoid the direct computation of derivatives, the quantities $\bm{\varphi}_f(t_k,\bm{\theta}_j)$, $\hat{\bm{\varphi}}_f(t_k,\bm{\theta}_j)$, and $y_f(t_k,\bm{\theta}_j)$ use filtered versions of the derivatives of the sampled input and output data. This filtering process is implemented in discrete-time based on a ZOH or FOH intersample behavior of the sampled signals. For the SRIVC method, these are given by
\begin{align}
	\bm{\varphi}_f(t_k,\bm{\theta}_j) &= \bigg[\frac{-p}{A_j(p)} y(t_k), \hspace{0.03cm} \dots,\hspace{0.03cm} \frac{-p^n}{A_j(p)} y(t_k), \notag \\
	\label{filteredregressor}
	&\hspace{1.2cm}\frac{1}{A_j(p)} u(t_k), \hspace{0.05cm}\dots,\hspace{0.05cm} \frac{p^m}{A_j(p)} u(t_k)\bigg]^\top, \\
	\hat{\bm{\varphi}}_{f}(t_k,\bm{\theta}_j) &=  \bigg[ \frac{-p B_j(p)}{A_j^2(p)} u(t_k),\hspace{0.05cm} \dots, \hspace{-0.05cm}\frac{-p^n B_j(p)}{A_j^2(p)} u(t_k), \notag \\
	\label{filteredinstrument_open}
	&\hspace{1.2cm} \frac{1}{A_j(p)} u(t_k), \hspace{0.05cm}\dots,\hspace{0.05cm} \frac{p^m}{A_j(p)} u(t_k)\bigg]^\top, \\
	\label{filteredoutput}
	y_f(t_k,\bm{\theta}_j) &= \frac{1}{A_j(p)}y(t_k), 
\end{align} 
where $A_j(p)$ and $B_j(p)$ are the polynomials of the model estimate at the $j$th iteration. For notation purposes, we might omit the dependence on $\bm{\theta}_j$ in $\hat{\bm{\varphi}}_f$, $\bm{\varphi}_f$ and $y_f$. We denote the model estimate at the $j$th iteration as $G_j(p)\hspace{-0.04cm}=\hspace{-0.04cm}B_j(p)/A_j(p)$, and its ZOH equivalent as $G_{\textnormal{d},j}(q)$. 

On the other hand, the CLSRIVC estimator uses the same filtered regressor and output as SRIVC but its filtered instrument vector needs to be chosen such that it is not correlated with the noise on the output in closed-loop settings. The noise-free plant input and output can be estimated from the sensitivity functions constructed using the previous estimate of the plant, which leads to
\begin{align}
	\hat{\bm{\varphi}}_{f}(t_k,\bm{\theta}_j)\hspace{-0.03cm} &= \hspace{-0.03cm}\bigg[\frac{-p B_j(p) S_{uo,j}(p)}{A_j^2(p)},\hspace{0.02cm}\dots,\hspace{0.02cm} \frac{-p^n B_j(p) S_{uo,j}(p)}{A_j^2(p)}, \notag \\
	\label{filteredinstrument_closed}
	&\hspace{0.5cm}\frac{ S_{uo,j}(p)}{A_j(p)}, \hspace{0.03cm}\dots,\hspace{0.03cm} \frac{p^m S_{uo,j}(p)}{A_j(p)}\bigg]^\top r(t_k),
\end{align}
where $S_{uo,j}(p)=C(p)/[1+G_j(p)C(p)]$ is the control sensitivity function estimate, computed using the $j$th iteration of the algorithm. One important aspect is that the CLSRIVC estimator, as introduced in \cite{garnier2008book}, assumes that a \textit{continuous-time} controller is known in advance for the implementation of the algorithm. Therefore, its standard form is only suitable for Setting 1. 

Throughout this paper we refer to the notion of \textit{generic} consistency \cite{soderstrom1984generic}, which, for the SRIVC and CLSRIVC estimators, is related to the generic non-singularity of the matrix being inverted in the algorithms. We introduce the concept of generic non-singularity in Definition \ref{def1}, and then we present the technical lemma that is used for proving this property in our context. For completeness, the definitions of consistency and generic consistency are presented in Definitions \ref{def:consistency} and \ref{def:genericconsistency}, respectively.
\begin{defn}[Generically non-singular matrix]
	\label{def1}
	Consider an $n\times n$ matrix $\mathbf{R}(\bm{\rho})$, which depends on a vector $\bm{\rho}$ belonging to an open subset $\Omega$ of the Euclidean space $\mathbb{R}^{n_\rho}$. Then, $\mathbf{R}$ is generically non-singular with respect to $\bm{\rho}\in \Omega$ if the set $\{\bm{\rho}\colon \bm{\rho}\in\Omega, \text{rank } \mathbf{R}(\bm{\rho})<n\}$ has Lebesgue measure zero in $\Omega$.
\end{defn}
\begin{lem}[Corollary of Lemma A2.3 in \cite{soderstrom1983instrumental}]
	\label{lemmageneric}
	Consider the matrix $\mathbf{R}$ and the set $\Omega$ given in Definition \ref{def1}. Assume that there is a vector $\bm{\rho}^*\in\Omega$ such that $\mathbf{R}(\bm{\rho}^*)$ is non-singular, and that the entries of $\mathbf{R}$ are analytic functions of every element in $\bm{\rho}\in\Omega$.	Then, $\mathbf{R}(\bm{\rho})$ is generically non-singular with respect to $\bm{\rho}\in\Omega$.
\end{lem}
\begin{pf*}{Proof.}
	By using Proposition 1 on the zero set of real analytic functions in \cite{mityagin2020zero}, the result follows directly from the proof of Lemma 1 of \cite{soderstrom1984generic}. \hspace*{\fill} \qed
\end{pf*}
\begin{defn}[Consistency]
	\label{def:consistency}
	An estimator $\hat{\bm{\theta}}_N$ with tuning parameters $\bm{\eta}$ is consistent for $\bm{\theta}^*$ and $\bm{\eta}$ if $\hat{\bm{\theta}}_N$ converges almost surely to $\bm{\theta}^*$.
\end{defn}
\begin{defn}[Generic consistency]
	\label{def:genericconsistency}
	An estimator $\hat{\bm{\theta}}_N$ with tuning parameters $\bm{\eta}$ is generically consistent with respect to $\bm{\rho}=[\bm{\theta}^{*\top},\bm{\eta}^\top]^\top \in \Omega$ if the set $\{\bm{\rho}\colon \bm{\rho}\in\Omega, \hat{\bm{\theta}}_N$ is not consistent for $\bm{\rho}\}$ has Lebesgue measure zero in $\Omega$.
\end{defn}
That is, a generic consistent estimator is one that is consistent for almost any choice in the tuning parameters of the method and for almost any system. The cases in which the estimator is not consistent, if they exist, are isolated.
\begin{rem}
The notion of consistency is closely linked to asymptotic unbiasedness. An estimator $\hat{\bm{\theta}}_N$ is consistent if it is asymptotically unbiased and its variance decays to zero \cite[p. 54]{lehmann1998theory}. Conversely, any estimator that is consistent is asymptotically unbiased if its mean square error is uniformly bounded. This fact follows from
	\begin{align}
		\mathbb{E}\{\|\hat{\bm{\theta}}_N\hspace{-0.04cm}-\hspace{-0.04cm}\bm{\theta}^*\|\}&\hspace{-0.04cm}<\hspace{-0.04cm}\epsilon \hspace{-0.04cm}+\hspace{-0.06cm} \sqrt{\mathbb{E}\{\|\hat{\bm{\theta}}_N\hspace{-0.04cm}-\hspace{-0.04cm}\bm{\theta}^*\|^2\}\mathbb{P}\{\|\hat{\bm{\theta}}_N\hspace{-0.04cm}-\hspace{-0.04cm}\bm{\theta}^*\|\geq \epsilon\}} \notag \\
		& \xrightarrow[]{N\to\infty,\epsilon\to 0} 0, \notag
	\end{align}
	where the second term arises from the Cauchy-Schwartz inequality applied to $\mathbb{E}\{\|\hat{\bm{\theta}}_N-\bm{\theta}^*\|\big|\|\hat{\bm{\theta}}_N-\bm{\theta}^*\|\geq \epsilon\}$.
\end{rem}

This paper provides a comprehensive analysis on the generic consistency of the SRIVC and CLSRIVC estimators in closed-loop for Settings 1 and 2. 

\section{Analysis of the SRIVC and CLSRIVC estimators for Setting 1}
\label{setup1}
In this section, we study the consistency of both refined instrumental variable methods for Setting 1. Most of our attention will be focused on the properties of the CLSRIVC estimator, as they have not been carefully addressed in the literature prior to this work. With regards to the SRIVC estimator, the following result follows directly from Corollary 3 of \cite{pan2020consistency}.

\begin{thm}\hspace{-0.2cm}\textbf{.}
	\label{thmsrivc_consistency}
	Consider the SRIVC estimator with filtered regressor \eqref{filteredregressor}, filtered instrument \eqref{filteredinstrument_open}, and filtered output \eqref{filteredoutput}, and suppose Assumptions \ref{assumption21} to \ref{assumption23} hold. Moreover, assume that $\mathbb{E}\{\hat{\boldsymbol\varphi}_f(t_k)\boldsymbol\varphi^\top_f(t_k)\}$ is generically non-singular with respect to the system and model parameters. The SRIVC estimator, provided it converges in iterations for all $N$ large enough, is not generically consistent nor asymptotically unbiased under Setting 1 when sampled data are available as measurements.
\end{thm}
\begin{pf*}{Proof.}
	The result follows as a direct consequence of Corollary 3 of \cite{pan2020consistency}, as it is known that a continuous-time controller $C(p)$ produces a smooth system input that cannot be exactly reconstructed from a ZOH device for the computation of the filtered regressor vector. \hspace*{\fill} \qed
\end{pf*}

The analysis for the CLSRIVC estimator is more intricate, and therefore we will state some additional technical assumptions that are related to the stability of the estimates, the reference excitation, and the sampling frequency. More precisely, the assumptions that we consider are the following:

\begin{assumption}	
	\label{assumption41}
	For all $j\geq 1$, the zeros of $A_j(p)$ have strictly negative real parts, $n \geq m$, with $A_j(p)$ and $B_j(p)$ being coprime. The model of the closed-loop system is also asymptotically stable, i.e., the zeros of $A_j(p)L(p)+B_j(p)F(p)$ have strictly negative real parts.
\end{assumption}
\begin{assumption}
	\label{assumption42}
	The external reference is persistently exciting of order no less than $2n+n_l^*$ if $n>m$ and no less than $2n+n_l^*+1$ if $n=m$.
\end{assumption}
\begin{assumption}	
	\label{assumption43}
	The sampling frequency is larger than twice the largest imaginary part of the zeros\footnote{Note that such assumption is not restrictive if the user has freedom to choose the sampling period, since for continuous-time system identification the sampling period can be chosen small without inherent problems of ill-conditioning in the continuous-time parameters \cite{garnier2014advantages}.} of $A_j(p)\big(A^*(p)L(p)+B^*(p)F(p)\big)$.
\end{assumption}

We shall define some signals before we present the main results. The filtered regressor vector can be written as $\bm{\varphi}_f(t_k) = \tilde{\bm{\varphi}}_f^r(t_k)+ \bm{\Delta}(t_k)-\textbf{v}_f(t_k),$ where $\mathbf{v}_f(t_k)$ only depends on the output noise, the noise-free, interpolation error-free regressor vector is given by
\begin{equation}
	\label{tildevarphi1}
	\tilde{\bm{\varphi}}_{\hspace{-0.02cm}f}^r\hspace{-0.02cm}(\hspace{-0.01cm}t_k\hspace{-0.01cm}) \hspace{-0.09cm}= \hspace{-0.12cm}\begin{bmatrix}
		\hspace{-0.05cm}\dfrac{p T_o^*\hspace{-0.03cm}(\hspace{-0.01cm}p\hspace{-0.01cm})}{-\hspace{-0.03cm}A_{\hspace{-0.02cm}j}\hspace{-0.03cm}(\hspace{-0.01cm}p\hspace{-0.01cm})}, & \hspace{-0.1cm} .\hspace{0.04cm}.\hspace{0.04cm}.\hspace{0.04cm}, & \hspace{-0.14cm} \dfrac{p^{\hspace{-0.02cm}n}\hspace{-0.04cm} T_{\hspace{-0.03cm}o}^*\hspace{-0.04cm}(\hspace{-0.01cm}p\hspace{-0.01cm})}{-\hspace{-0.03cm}A_{\hspace{-0.02cm}j}(\hspace{-0.01cm}p\hspace{-0.01cm})}, & \hspace{-0.14cm} \dfrac{S_{\hspace{-0.035cm}uo}^*\hspace{-0.02cm}(\hspace{-0.01cm}p\hspace{-0.01cm})}{A_j\hspace{-0.02cm}(\hspace{-0.01cm}p\hspace{-0.01cm})}, & \hspace{-0.1cm} .\hspace{0.04cm}.\hspace{0.04cm}.\hspace{0.04cm}, & \hspace{-0.14cm} \dfrac{p^{\hspace{-0.02cm}m} \hspace{-0.04cm}S_{\hspace{-0.035cm}uo}^*\hspace{-0.03cm}(\hspace{-0.01cm}p\hspace{-0.01cm})}{A_j(p)}
	\end{bmatrix}^{\hspace{-0.08cm}\top} \hspace{-0.17cm}r\hspace{-0.02cm}(\hspace{-0.015cm}t_k\hspace{-0.015cm}),
\end{equation}
and $\bm{\Delta}(t_k)$ is a vector that contains the interpolation errors that arise from constructing the filtered disturbance-free derivatives of the output. The entries of $\bm{\Delta}(t_k)$ are given by the difference between the noise-free version of the regressor vector and $\tilde{\bm{\varphi}}_f^r(t_k)$ in \eqref{tildevarphi1}, i.e.,
\begin{equation}
	\label{delta1}
	\bm{\Delta}_i\hspace{-0.02cm}(\hspace{-0.01cm}t_k\hspace{-0.01cm}) \hspace{-0.1cm}=\hspace{-0.13cm} \begin{cases}
		\hspace{-0.11cm}\dfrac{p^i \hspace{-0.02cm}T_o^*\hspace{-0.035cm}(\hspace{-0.01cm}p\hspace{-0.01cm})}{A_j(p)}r(\hspace{-0.01cm}t_k\hspace{-0.01cm}) \hspace{-0.08cm}- \hspace{-0.1cm}\dfrac{p^i}{A_{\hspace{-0.02cm}j}\hspace{-0.02cm}(\hspace{-0.01cm}p\hspace{-0.01cm})}\hspace{-0.05cm}\left\{\hspace{-0.02cm}T_o^*\hspace{-0.03cm}(\hspace{-0.01cm}p\hspace{-0.01cm})r(\hspace{-0.01cm}t\hspace{-0.01cm})\hspace{-0.02cm}\right\}_{\hspace{-0.02cm}t=t_k}\hspace{-0.1cm}, & \hspace{-0.29cm}i\hspace{-0.09cm}=\hspace{-0.09cm}1,.\hspace{0.04cm}.\hspace{0.04cm}.\hspace{0.04cm},n, \\
		0, & \hspace{-0.3cm} \textnormal{otherwise}.
	\end{cases}
\end{equation}
In Lemma \ref{lemmainv} we show that the modified normal matrix being inverted in each iteration of the CLSRIVC method is generically non-singular when $N$ tends to infinity. This property is essential for establishing asymptotic properties of the converging point of the method, since it ensures that the converging point is well-defined in all but possibly some rare and isolated cases. In the sequel, we denote $\|\cdot\|_2$ and $\sigma_{\textnormal{min}}(\cdot)$ as the 2-norm and minimum singular value of a matrix, respectively.

\begin{lem}
	\label{lemmainv}
	Consider the CLSRIVC estimator with iterations \eqref{iterations}, with filtered regressor vector \eqref{filteredregressor}, filtered instrument vector \eqref{filteredinstrument_closed} and filtered output \eqref{filteredoutput}. Suppose Assumptions~\ref{assumption21} to \ref{assumption23} and \ref{assumption41} to \ref{assumption43} hold, and that the condition
	\begin{equation}
		\label{conditioninv}
		\big\|\mathbb{E}\hspace{-0.05cm}\left\{\hspace{-0.03cm}\hat{\boldsymbol{\varphi}}_f(t_k)\boldsymbol{\Delta}^{\hspace{-0.04cm}\top}\hspace{-0.04cm}(t_k)\hspace{-0.02cm}\right\}\hspace{-0.07cm}\big\|_2\hspace{-0.06cm}<\hspace{-0.04cm}\sigma_{\textnormal{min}}\hspace{-0.07cm}\left(\hspace{-0.02cm}\mathbb{E}\big\{\hspace{-0.02cm}\hat{\boldsymbol{\varphi}}_f(t_k){{}{\tilde{\boldsymbol{\varphi}}_f^r}}^{\hspace{-0.04cm}\top}(t_k)\big\}\hspace{-0.03cm}\right) \hspace{-0.2cm}
	\end{equation}
	is satisfied, where $\tilde{\boldsymbol{\varphi}}_f^r(t_k)$ and $\boldsymbol{\Delta}(t_k)$ are given by~\eqref{tildevarphi1} and \eqref{delta1} respectively. Then, the modified normal matrix $\mathbb{E}\{\hat{\boldsymbol\varphi}_f(t_k)\boldsymbol\varphi^\top_f(t_k)\}$ of the CLSRIVC method is generically non-singular with respect to the system denominator and model parameters.  
\end{lem}
\begin{pf*}{Proof.}
	The vectors $\hat{\boldsymbol\varphi}_{f}(t_k)$ and $\tilde{\boldsymbol\varphi}_f^r(t_k)$ can be written~as
	\begin{align}
		\label{equivalent1}
		\hat{\boldsymbol\varphi}_{f}(t_k) &= \mathbf{S}(-B_j,A_j)\frac{S_{uo,j}(p)}{A_j^2(p)} \mathbf{r}_{n+m}(t_k), \\
		\label{equivalent2}
		\tilde{\boldsymbol\varphi}_f^r(t_k) &= \mathbf{S}(-B^*,A^*) \frac{S_{uo}^*(p)}{A_j(p)A^{*}(p)}\mathbf{r}_{n+m}(t_k)
	\end{align}
	respectively, where $\mathbf{S}(-B_j,A_j)$ and $\mathbf{S}(-B^*,A^*)$ are $(n+m+1)\times(n+m+1)$ Sylvester matrices constructed using the coefficients of the respective $B$ and $A$ polynomials. These matrices are non-singular under Assumptions \ref{assumption41} and \ref{assumption21} due to Lemma A3.1 of \cite{soderstrom1983instrumental}. The vector $\mathbf{r}_{n+m}(t_k)$ contains the derivatives of the reference signal, that is,
	\begin{equation}
		\label{eq:rd}
		\mathbf{r}_{n+m}(t_k) = \begin{bmatrix}
			p^{n+m}, & p^{n+m-1}, & \dots, & 1
		\end{bmatrix}^\top r(t_k).
	\end{equation}
	As $N\rightarrow\infty$, the ergodicity lemma in \cite[Lemma 3.1]{soderstrom1975ergodicity} permits expressing the modified normal matrix of the CLSRIVC estimator as
	\begin{align}	
		\mathbb{E}&\left\{\hat{\boldsymbol\varphi}_{f}(t_k)\boldsymbol\varphi^\top_f(t_k)\right\} =\mathbf{S}(-B_j,A_j)\boldsymbol\Phi\mathbf{S}^{\top}(-B^*,A^*)\notag \\
		\label{eq:final_normal}
		&\hspace{0.5cm}+\mathbb{E}\left\{\hat{\boldsymbol\varphi}_{f}(t_k) \boldsymbol\Delta^{\top}(t_k)\right\} +\mathbb{E}\left\{\hat{\boldsymbol\varphi}_{f}(t_k) \mathbf{v}_f^\top(t_k)\right\},
	\end{align}
	where
	\begin{equation}
		\label{eq:Phi}
		\boldsymbol\Phi \hspace{-0.06cm}:= \hspace{-0.05cm}\mathbb{E}\hspace{-0.02cm}\bigg\{\frac{S_{uo,j}\hspace{-0.03cm}(p)}{A_j^2(p)} {\mathbf{r}_{n\hspace{-0.02cm}+\hspace{-0.02cm}m}}\hspace{-0.03cm}(t_k)\frac{S_{uo}^*(p)}{A_j(p)A^*(p)} \mathbf{r}_{n\hspace{-0.02cm}+\hspace{-0.02cm}m}^\top\hspace{-0.03cm}(t_k)\hspace{-0.03cm}\bigg\}\hspace{-0.04cm}. \hspace{-0.2cm}
	\end{equation}
	By the same arguments found in the proof of Lemma 7 of \cite{pan2020consistency}, the last expectation in~\eqref{eq:final_normal} is equal to zero under Assumption~\ref{assumption22}. Thus, Theorem 5.1 of \cite{dahleh2002lectures} ensures the (generic) non-singularity of $\mathbb{E}\{\hat{\bm{\varphi}}_f(t_k){{}\bm{\varphi}}_f^\top(t_k)\}$ provided that \eqref{conditioninv} holds and that $\bm{\Phi}$ is generically non-singular. The generic non-singularity of $\bm{\Phi}$ follows from the arguments below. Let the set of parameters that describe $A_j(p)$ and $B_j(p)$ be defined as $
			\Omega \hspace{-0.07cm}=\hspace{-0.09cm} \big\{\bm{\theta}_j \hspace{-0.03cm}\in \hspace{-0.02cm}\mathbb{R}^{n+m+1} \hspace{-0.02cm}\colon \hspace{-0.02cm}A_j(p) \; \text{is a stable polynomial}\hspace{0.03cm}\big\}\hspace{-0.02cm}.$ Define $\boldsymbol\Phi^*$ as the matrix $\boldsymbol{\Phi}$ in \eqref{eq:Phi} with $A_j(p)=A^*(p)$ and $B_j(p)=B^*(p)$. Note that all the transfer functions that form the elements of $\boldsymbol\Phi^*$ are proper due to the second condition in Assumption \ref{assumption23}. The denominator polynomial of $S_{uo}^*(p)/{{}A^*}^2(p)$ has a degree of $n+\max(n+n_l^*,m+n_f^*)=2n+n_l^*$. Then, by the same procedure as in the proof of Lemma 7 in \cite{pan2020consistency}, $\boldsymbol\Phi^*$ can be shown to be positive definite if the reference signal is persistently exciting of order $2n+n_l^*$ in case $n>m$, or $2n+n_l^*+1$ in case $n=m$. Next, by leveraging the frequency domain description of covariance expressions, an arbitrary entry of $\boldsymbol\Phi$ in~\eqref{eq:Phi} can be written as
	\begin{align*}
		\boldsymbol\Phi_{il} \hspace{-0.06cm} &= \hspace{-0.05cm}\mathbb{E}\hspace{-0.06cm}\left\{\hspace{-0.04cm}\frac{p^{n+m+1-i} S_{uo,j}(p)}{A_j^2(p)}r(t_k)
		\frac{p^{n+m+1-l}S_{uo}^*(p)}{A_j(p)A^*(p)}r(t_k)\right\}\\
		&\hspace{-0.55cm}=\hspace{-0.17cm}\int_{\hspace{-0.06cm}-\hspace{-0.02cm}\pi}^{\pi}\hspace{-0.12cm}\frac{C_1(e^{i\omega})C_2(e^{-i\omega})/(2\pi)}{|\hspace{-0.02cm}A_{\textnormal{d}\hspace{-0.01cm},\hspace{-0.02cm}j}\hspace{-0.03cm}(\hspace{-0.025cm}e^{i\omega}\hspace{-0.025cm})\hspace{-0.02cm}|^2 \hspace{-0.04cm}A_{\mathrm{d}\hspace{-0.01cm},\hspace{-0.02cm}j}\hspace{-0.03cm}(\hspace{-0.025cm}e^{i\omega}\hspace{-0.025cm})Q_{\hspace{-0.025cm}\textnormal{d},\hspace{-0.02cm}j}\hspace{-0.04cm}(\hspace{-0.025cm}e^{i\omega}\hspace{-0.025cm})\hspace{-0.02cm}A_\textnormal{d}^*\hspace{-0.03cm}(\hspace{-0.025cm}e^{\hspace{-0.02cm}-i\omega}\hspace{-0.02cm})Q_{\hspace{-0.02cm}\textnormal{d}}^*\hspace{-0.04cm}(\hspace{-0.025cm}e^{\hspace{-0.02cm}-i\omega}\hspace{-0.025cm})}
	 \mathrm{d}\hspace{-0.02cm}F_{\hspace{-0.03cm}r}\hspace{-0.03cm}(\hspace{-0.02cm}\omega\hspace{-0.02cm})\hspace{-0.01cm},
	\end{align*}
	where $F_r(\omega)$ is the spectral distribution of the reference signal, and $A_{\textnormal{d},j}, Q_{\textnormal{d},j}, A_\textnormal{d}^*$ and $Q_\textnormal{d}^*$ are the denominator polynomials of the respective ZOH equivalents. Following the same steps as in the proof of Lemma 9 in \cite{pan2020consistency} shows that the entries of $\boldsymbol\Phi$ are analytic functions in $\Omega$. Thus, invoking Lemma \ref{lemmageneric} leads to the conclusion that the modified normal matrix $\mathbb{E}\{\hat{\boldsymbol\varphi}_f(t_k)\boldsymbol\varphi^\top_f(t_k)\}$ is generically non-singular. \hspace*{\fill} \qed
\end{pf*}
Next, we examine the consistency of the CLSRIVC estimator in Theorem~\ref{thmclsrivc_consistency}.
\begin{thm}\hspace{-0.2cm}\textbf{.}
	\label{thmclsrivc_consistency}
	Consider the CLSRIVC estimator with iterations \eqref{iterations}, filtered regressor vector \eqref{filteredregressor}, filtered instrument vector \eqref{filteredinstrument_closed} and filtered output \eqref{filteredoutput}. Suppose Assumptions \ref{assumption21} to \ref{assumption43} hold, and that the condition in \eqref{conditioninv} is satisfied. Then, provided the CLSRIVC estimator converges in iterations for all $N$ sufficiently large, the estimator is not generically consistent nor asymptotically unbiased under Setting 1 when only sampled data are available as measurements.
\end{thm}

\begin{pf*}{Proof.}
	If we denote $\bar{\bm{\theta}}^N$ as the converging point of the CLSRIVC estimator (with associated transfer function model $\bar{G}_N(p)=\bar{B}_N(p)/\bar{A}_N(p)$), it must satisfy
	\begin{equation}
		\bar{\bm{\theta}}^{\hspace{-0.02cm}N}\hspace{-0.16cm}=\hspace{-0.13cm} \left[\sum_{k=1}^N \hspace{-0.1cm}\hat{\bm{\varphi}}_{\hspace{-0.025cm}f}\hspace{-0.02cm}(\hspace{-0.015cm}t_k\hspace{-0.02cm},\hspace{-0.03cm}\bar{\bm{\theta}}^{\hspace{-0.02cm}N}\hspace{-0.035cm}) \bm{\varphi}_{\hspace{-0.025cm}f}^{\hspace{-0.02cm}\top}\hspace{-0.06cm}(\hspace{-0.015cm}t_k\hspace{-0.02cm},\hspace{-0.03cm}\bar{\bm{\theta}}^{\hspace{-0.02cm}N}\hspace{-0.035cm})\hspace{-0.03cm}\right]^{\hspace{-0.08cm}-\hspace{-0.02cm}1}\hspace{-0.13cm}\left[\sum_{k=1}^N  \hspace{-0.09cm}\hat{\bm{\varphi}}_{\hspace{-0.025cm}f}\hspace{-0.02cm}(\hspace{-0.015cm}t_k,\hspace{-0.03cm}\bar{\bm{\theta}}^{\hspace{-0.02cm}N}\hspace{-0.035cm}) y_{\hspace{-0.025cm}f}\hspace{-0.02cm}(\hspace{-0.015cm}t_k,\hspace{-0.03cm}\bar{\bm{\theta}}^{\hspace{-0.02cm}N}\hspace{-0.035cm})\hspace{-0.02cm}\right]\hspace{-0.1cm}. \notag 
	\end{equation}
	The ergodicity lemmas in \cite[Lemma 3.1]{soderstrom1975ergodicity} and \cite[Lemma A4.3]{soderstrom1983instrumental} permit writing the equation above when $N$ tends to infinity as
	\begin{align}
		\mathbb{E}&\left\{\hat{\boldsymbol\varphi}_f(t_k,\bar{\boldsymbol\theta})	
		\boldsymbol\varphi^\top_f(t_k,\bar{\boldsymbol\theta})\right\}^{-1} \notag \\
		&\hspace{0.1cm}\times \mathbb{E}\big\{\hat{\boldsymbol\varphi}_f(t_k,\bar{\boldsymbol\theta}) 
		\left(y_f(t_k,\bar{\boldsymbol\theta}) - \boldsymbol\varphi^\top_f(t_k,\bar{\boldsymbol\theta})\bar{\boldsymbol\theta}\right)\hspace{-0.07cm}\big\} = \mathbf{0}, \notag
	\end{align}
	where $\bar{\bm{\theta}} = \lim_{N\to \infty} \bar{\bm{\theta}}^N$, with an associated transfer function model $\bar{G}(p)=\bar{B}(p)/\bar{A}(p)$. The non-singularity of the modified normal matrix implies that
	\begin{align}
		&\mathbb{E}\big\{\hat{\boldsymbol\varphi}_f(t_k,\bar{\boldsymbol\theta}) 
		\left(y_f(t_k,\bar{\boldsymbol\theta}) - \boldsymbol\varphi^\top_f(t_k,\bar{\boldsymbol\theta})\bar{\boldsymbol\theta}\right)\hspace{-0.05cm}\big\} 
		= \mathbf{0}	\notag \\
		&\iff\mathbb{E}\big\{\hat{\boldsymbol\varphi}_f(t_k,\bar{\boldsymbol\theta}) 
		\big(\hspace{-0.05cm}\left\{G^*(p)u(t)\right\}_{t=t_k} -\bar{G}(p)u(t_k) \big)\big\} \notag \\
		&\hspace{0.95cm}+ \mathbb{E}\big\{\hat{\boldsymbol\varphi}_f(t_k,\bar{\boldsymbol\theta}) v(t_k)\big\} 
		= \mathbf{0}. \notag
	\end{align}
	Since the additive noise on the output is independent of the reference signal under Assumption \ref{assumption22}, 
	it follows that $\mathbb{E}\{\hat{\boldsymbol\varphi}_f(t_k,\bar{\boldsymbol\theta}) v(t_k)\} = \mathbf{0}$. Hence, the converging point of the CLSRIVC estimator satisfies
	\begin{align}
		\label{untilthispoint}
		&\mathbb{E}\left\{\hat{\boldsymbol\varphi}_f(t_k,\bar{\boldsymbol\theta}) 
		\left(\{G^*(p)u(t)\}_{t=t_k}\hspace{-0.05cm} - \hspace{-0.05cm}\bar{G}(p)u(t_k) \right)\hspace{-0.02cm}\right\} = \mathbf{0} \\
		\label{eq:epsilon}
		&\iff\mathbb{E}\Big\{\hat{\boldsymbol\varphi}_f(t_k,\bar{\boldsymbol\theta}) \varepsilon(t_k,\bar{\boldsymbol\theta})\Big\} = \mathbf{0},
	\end{align}
	where $\varepsilon(\hspace{-0.02cm}t_k,\hspace{-0.03cm}\bar{\boldsymbol\theta}\hspace{-0.02cm})\hspace{-0.1cm}:\hspace{-0.02cm}=\hspace{-0.1cm}\{\hspace{-0.025cm}G^*\hspace{-0.04cm}(p)S_{\hspace{-0.025cm}uo}^*\hspace{-0.03cm}(\hspace{-0.01cm}p\hspace{-0.01cm})r\hspace{-0.02cm}(\hspace{-0.01cm}t\hspace{-0.01cm})\}_{\hspace{-0.02cm}t=t_k} \hspace{-0.1cm}-\hspace{-0.02cm}\bar{G}\hspace{-0.025cm}(\hspace{-0.02cm}p\hspace{-0.02cm})\hspace{-0.07cm}\left\{\hspace{-0.03cm}S_{\hspace{-0.02cm}uo}^*\hspace{-0.03cm}(\hspace{-0.01cm}p\hspace{-0.01cm})r(\hspace{-0.01cm}t\hspace{-0.01cm})\hspace{-0.03cm}\right\}_{t=t_k}\hspace{-0.05cm}$. Now, we introduce a reference-dependent term
	\begin{equation}
		\label{varepsilonreference}
		\varepsilon_r(t_k,\hspace{-0.02cm}\bar{\bm{\theta}}) \hspace{-0.08cm}:= \hspace{-0.09cm}\left\{\hspace{-0.04cm}\bar{G}\hspace{-0.02cm}(p)S_{uo}^*(p)r(t)\right\}_{\hspace{-0.03cm}t=t_k}
		\hspace{-0.13cm}-\hspace{-0.02cm}\bar{G}(p)\hspace{-0.06cm}\left\{\hspace{-0.02cm}S_{uo}^*(p)r(t)\right\}_{t=t_k}\hspace{-0.09cm}.
	\end{equation}
	Then, $\varepsilon(t_k,\bar{\boldsymbol\theta})$ in terms of $\varepsilon_r(t_k,\bar{\boldsymbol\theta})$ becomes
	\begin{align*}
		\varepsilon(t_k,\bar{\boldsymbol\theta}) \hspace{-0.06cm}&=\hspace{-0.06cm} G^*(p)S_{uo}^*(p)r(t_k)\hspace{-0.04cm}-\hspace{-0.04cm}\bar{G}(p)S_{uo}^*(p)r(t_k) \hspace{-0.05cm}+\hspace{-0.05cm} \varepsilon_r(t_k,\bar{\bm{\theta}}) \notag \\
		&=\hspace{-0.06cm} \frac{S_{uo}^*(p)}{\bar{A}(p)A^*(p)}\mathbf{r}_{n+m}^\top(t_k)\mathbf{h} \hspace{-0.03cm}+\hspace{-0.03cm} \varepsilon_r(t_k,\bar{\bm{\theta}}), \notag
	\end{align*}
	where $\mathbf{r}_{n+m}(t_k)$ is defined in \eqref{eq:rd} and $\mathbf{h}$ is a vector of dimension $(n+m+1)$ containing the coefficients of $B^*(p)\bar{A}(p)-\bar{B}(p)A^*(p)$ in descending order of degree. Using the fact that the filtered instrument vector can be written in terms of $\mathbf{r}_{n+m}(t_k)$ as in~\eqref{equivalent1}, with $\mathbf{S}(\hspace{-0.03cm}-\bar{B},\hspace{-0.03cm}\bar{A})$ being non-singular given Assumption \ref{assumption41}, the condition in~\eqref{eq:epsilon} can be expressed as
	\begin{align}
		&\underbrace{\mathbb{E}\left\{\frac{\bar{S}_{uo}(p)}{\bar{A}^2(p)}\mathbf{r}_{n+m}(t_k)
			\frac{S_{uo}^*(p)}{\bar{A}(p)A^*(p)}\mathbf{r}_{n+m}^\top(t_k)\right\}}_{=:\bar{\boldsymbol\Phi}}\mathbf{h} \notag \\
		&\hspace{1.3cm}+\underbrace{\mathbb{E}\left\{\frac{\bar{S}_{uo}(p)}{\bar{A}^2(p)}\mathbf{r}_{n+m}(t_k)
			\varepsilon_r(t_k,\bar{\bm{\theta}})\right\}}_{=:\bar{\boldsymbol\Psi}} =\mathbf{0}. \notag
	\end{align}
	We can show that $\bar{\boldsymbol\Phi}$ is generically non-singular by the same reasoning as in Lemma \ref{lemmainv}. Note that $\bar{\boldsymbol\Psi}$ is reference-dependent and is in general not equal to zero due to $\varepsilon_r(t_k,\bar{\bm{\theta}}) \neq 0$. Since the matrix $\mathbf{S}(-\bar{B},\bar{A})$ is non-singular under Assumption~\ref{assumption41}, we have
	\begin{equation}
		\label{hcomputation}
		\mathbf{h} = -\bar{\boldsymbol\Phi}^{-1}\bar{\boldsymbol\Psi} \neq \mathbf{0}.
	\end{equation}
	This in turn implies that $\bar{B}(p)/\bar{A}(p) \neq B^*(p)/A^*(p)$, i.e., the unique converging point does not correspond to the true parameter vector. We then conclude that the CLSRIVC estimator is generically not consistent nor asymptotically unbiased. \hspace*{\fill} \qed
\end{pf*}

\subsection{Reducing the bias of refined instrumental variable methods in closed-loop via oversampling}
Just as for the SRIVC estimator in the open-loop case, the lack of consistency of the CLSRIVC estimator is due to the unknown intersample behavior of the plant input $u(t)$. However, if the intersample behavior of the input were to be fully known and the filtered regressor vector of the CLSRIVC algorithm were to be computed as
\begin{align}
	\bm{\varphi}_{f}&(t_k,\bm{\theta}_j) = \bigg[\frac{-p}{A_j(p)} y(t_k), \hspace{0.05cm} \dots, \hspace{0.05cm} \frac{-p^n}{A_j(p)} y(t_k), \notag \\
	\label{bestfilteredregressor}
	&\left\{\frac{1}{A_j(p)} u(t) \right\}_{t=t_k}, \hspace{0.03cm}\dots,\hspace{0.03cm} \left\{\frac{p^m}{A_j(p)} u(t)\right\}_{t=t_k}\bigg]^{\top},
\end{align}
then the converging point would satisfy (cf. \eqref{eq:epsilon})
\begin{equation}
	\mathbb{E}\Big\{\hat{\boldsymbol\varphi}_f(t_k,\bar{\boldsymbol\theta}) 
	\left(\left\{[G^*(p)-\bar{G}(p)]S_{uo}^*(p)r(t)\right\}_{t=t_k}\right)\hspace{-0.06cm}\Big\} = \mathbf{0}. \notag
\end{equation}
Thus, $\mathbf{h}=\mathbf{0}$ and the CLSRIVC estimator would be generically consistent for this case. The uncertainty related to the intersample behavior of the input can be mitigated by choosing a faster sampling rate; however, from a theoretical standpoint, this can only alleviate the bias to a certain degree. If the input can be sampled at a faster rate than the output, then computing a more precise approximation of \eqref{bestfilteredregressor} constitutes a useful refinement of the CLSRIVC estimator, which has been studied as the CLSRIVC-os estimator in \cite{gonzalez2021srivc}. 

More precisely, oversampling procedures can reduce the norm of the reference-dependent term $\varepsilon_r(t_k,\bar{\bm{\theta}})$ that contributes to the bias of the CLSRIVC estimator via $\bar{\bm{\Psi}}$ in \eqref{hcomputation}. To analyze how over-sampling affects $\varepsilon_r(t_k,\bar{\bm{\theta}})$, we define the filtered regressor vector \eqref{bestfilteredregressor} but with $\tilde{u}(t)$ instead of $u(t)$, where $\tilde{u}(t)$ is the input signal after passing through a ZOH device with sampling instants $\mathcal{T}:=\{\tau_l\}_{l=1}^{M_k}$, with $M_k\gg k$, $\tau_{1}=t_1, \tau_{M_k}=t_k$ and $t_s\in \mathcal{T}$ for $s=1,2,\dots,k$. Furthermore, let $r_S(t):= S_{uo}^*(p)r(t)$. We have that $\varepsilon_r(t_k,\bar{\bm{\theta}}) = \{\bar{G}(p)\Delta r_S(t)\}_{t=t_k}$, where $\Delta r_S(t) = r_S(t)-r_S(\tau_l)$ when $t\in[\tau_l,\tau_{l+1})$, for $l=1,\dots,M_k-1$. Thus, after standard operations,
\begin{align}
	|\varepsilon_r(t_k,\bar{\bm{\theta}})|&\hspace{-0.04cm}\leq \|\bar{g}\|_1 \max_{t_1\leq t\leq t_k} |\Delta r_S(t)| \notag \\
	\label{rhs}
	&\hspace{-1.2cm}\leq \hspace{-0.05cm}\|\bar{g}\|_{\hspace{-0.02cm}1} \hspace{-0.1cm}\max_{l\in\{1,\dots,M_k-1\}}\hspace{-0.23cm}(\tau_{l+1}\hspace{-0.08cm}-\hspace{-0.06cm}\tau_l) \hspace{-0.35cm}\sup_{\substack{t_1<t<t_k \\ t\neq \tau_l, l=1,\dots,M_k}} \hspace{-0.15cm} \left|\frac{\textnormal{d}r_S(t)}{\textnormal{d}t}\right|,  
\end{align}
where $\bar{g}(t)$ is the impulse response of the system $\bar{G}(p)$. The supremum on the right-hand side of \eqref{rhs} depends on the reference as well as on the frequency response of the system $S_{uo}^*(p)$, and is bounded in the domain of interest if the reference is bounded and $S_{uo}^*(p)$ is asymptotically stable. In such case, the right-hand side will be negligible if $\tau_{l+1}-\tau_l$ is sufficiently small for all $l$. Note that the derivation above also indicates that in order to achieve a significant reduction in $|\varepsilon_r(t_k,\bar{\bm{\theta}})|$, it is convenient to sample more often when $r_S(t)$ changes more rapidly.

\begin{rem}
	Note that the filtered regressor vector in \eqref{bestfilteredregressor} has also been proposed in \cite{gonzalez2020consistent} for an extended version of the SRIVC estimator for multisine inputs, and has been considered for arbitrary inputs in \cite{gonzalez2021srivc}. 
\end{rem}

In summary, we have proven that the SRIVC and CLSRIVC estimators are not consistent when there is a continuous-time controller in the feedback loop and only sampled data are available as measurements. However, these estimators may benefit from a more sophisticated filtering that makes a more adequate assumption on the intersample behavior of the input signal $u(t)$.

\section{Analysis of the SRIVC and CLSRIVC estimators for Setting 2}
\label{setup2}
We now analyze the statistical properties of the refined instrumental variable methods when a discrete-time controller is present in the loop instead of a continuous-time one. This means that the intersample behavior of the system input is known to be a ZOH\footnote{Results when the intersample behavior of the input signal is FOH can be readily obtained, with only some minor modifications on the persistence of excitation condition of the external reference (see, e.g., Corollary 2 of \cite{pan2020consistency}).}, which is beneficial for the implementation and performance of the algorithms.

The analysis below relies on leveraging the following filtered versions of the external reference and disturbance:
\begin{equation}
	\label{tilderv}
	\tilde{r}(\hspace{-0.01cm}t_k\hspace{-0.01cm})\hspace{-0.09cm}:=\hspace{-0.08cm} \frac{C_\textnormal{d}(q)}{1\hspace{-0.07cm}+\hspace{-0.06cm}G_\textnormal{d}^*\hspace{-0.03cm}(\hspace{-0.01cm}q\hspace{-0.01cm})C_{\hspace{-0.02cm}\textnormal{d}}\hspace{-0.02cm}(\hspace{-0.01cm}q\hspace{-0.01cm})} r(\hspace{-0.01cm}t_k\hspace{-0.01cm}), \hspace{0.06cm} \tilde{v}(t_k)\hspace{-0.09cm}:=\hspace{-0.08cm} \frac{C_\textnormal{d}(q)}{1\hspace{-0.07cm}+\hspace{-0.06cm}G_\textnormal{d}^*\hspace{-0.03cm}(\hspace{-0.01cm}q\hspace{-0.01cm})C_{\hspace{-0.02cm}\textnormal{d}}\hspace{-0.02cm}(\hspace{-0.01cm}q\hspace{-0.01cm})} v(\hspace{-0.01cm}t_k\hspace{-0.01cm}).
\end{equation}
The assumptions we consider are the following:
\begin{assumption}
	\label{assumption51}	
	The transfer function $G_\textnormal{d}^*(q)C_\textnormal{d}(q)$ is strictly proper.
\end{assumption}
\begin{assumption}
	\label{assumption52}
	For all $j\geq 1$, all the zeros of $A_j(p)$ have strictly negative real parts, with $A_j(p)$ and $B_j(p)$ being coprime.
\end{assumption}
\begin{assumption}
	\label{assumption53}
	The external reference is persistently exciting of order no less than $2n+n_f^*$ if $n>m$ and no less than $2n+n_f^*+1$ if $n=m$.
\end{assumption}
\begin{assumption}
	\label{assumption54}	
	The sampling frequency is more than twice the largest positive imaginary part of the zeros of $A_j(p)A^*(p)$.
\end{assumption}
We must decompose the filtered regressor and instrument vectors similarly to the derivation in the previous section. For the sake of simplifying the notation, we will repeat the notation used in Section \ref{setup1} despite some vectors having different definitions. The filtered regressor vector can now be written as
\begin{equation}
	\label{decompositionregressor2}
	\bm{\varphi}_f(t_k) = \tilde{\bm{\varphi}}_f^r(t_k)+ \bm{\Delta}(t_k)-\textbf{v}_f(t_k),
\end{equation}
where
\begin{align}
	&\tilde{\bm{\varphi}}_f^r(t_k) = \bigg[
		\dfrac{-p B^*(p)}{A_j(p)A^*(p)}\tilde{r}(t_k), \dots, \dfrac{-p^n B^*(p)}{A_j(p) A^*(p)}\tilde{r}(t_k), \notag \\
		\label{tildevarphir}
	& \hspace{2cm}	 \dfrac{1}{A_j(p)}\tilde{r}(t_k), \dots, \dfrac{p^m}{A_j(p)}\tilde{r}(t_k)
	\bigg]^{\top}.
\end{align}
The entries of $\bm{\Delta}(t_k)$ are given by
\begin{equation}
	\label{delta}
	\bm{\Delta}_{i}\hspace{-0.02cm}(\hspace{-0.01cm}t_k\hspace{-0.01cm}) \hspace{-0.07cm}=\hspace{-0.08cm} \begin{cases}
		\hspace{-0.06cm}\frac{p^i G^{\hspace{-0.03cm}*}(p)}{A_j(p)}\tilde{r}(t_k) \hspace{-0.06cm}-\hspace{-0.06cm} \frac{p^i}{A_j(p)}\hspace{-0.03cm}\big[G_\textnormal{d}^*(q)\tilde{r}(t_k)\big], &\hspace{-0.15cm} i\hspace{-0.06cm}=\hspace{-0.05cm}1,\hspace{-0.02cm}\dots,\hspace{-0.02cm}n, \\
		0, & \textnormal{otherwise},
	\end{cases}
\end{equation}
and the contribution of the disturbance $v(t_k)$ in the filtered regressor is decomposed into two parts $\textbf{v}_f(t_k)=\textbf{v}_{f1}(t_k)+\textbf{v}_{f2}(t_k)$, which are given by
\begin{align}
	\label{varphiv1}
	\textbf{v}_{\hspace{-0.03cm}f1}\hspace{-0.04cm}(\hspace{-0.01cm}t_k\hspace{-0.01cm}) \hspace{-0.06cm} &= \hspace{-0.1cm} \begin{bmatrix}
		\dfrac{p}{A_{\hspace{-0.02cm}j}(\hspace{-0.01cm}p\hspace{-0.01cm})}, & \hspace{-0.1cm} \dots, & \hspace{-0.1cm} \dfrac{p^n}{A_{\hspace{-0.02cm}j}(\hspace{-0.01cm}p\hspace{-0.01cm})}, & \hspace{-0.07cm} 0, & \hspace{-0.05cm}\dots, & \hspace{-0.05cm} 0
	\end{bmatrix}^{\hspace{-0.07cm}\top} \hspace{-0.1cm}S_o^*(q)v(t_k), \hspace{-0.05cm}\\
	\label{varphiv2}
	\textbf{v}_{\hspace{-0.03cm}f2}(\hspace{-0.01cm}t_k\hspace{-0.01cm}) \hspace{-0.07cm} &= \hspace{-0.08cm} \begin{bmatrix}
		0, & \hspace{-0.1cm} \dots, & \hspace{-0.06cm} 0, & \hspace{-0.09cm} \dfrac{1}{A_{\hspace{-0.02cm}j}(\hspace{-0.01cm}p\hspace{-0.01cm})}\tilde{v}(\hspace{-0.01cm}t_k\hspace{-0.01cm}), & \hspace{-0.1cm}\dots, & \hspace{-0.09cm} \dfrac{p^m}{A_{\hspace{-0.02cm}j}(\hspace{-0.01cm}p\hspace{-0.01cm})}\tilde{v}(t_k)
	\end{bmatrix}^{\hspace{-0.06cm}\top}\hspace{-0.06cm}.
\end{align}
On the other hand, the filtered instrument vector can be decomposed as $\hat{\bm{\varphi}}_f(t_k) = \hat{\bm{\varphi}}_f^r(t_k) - \hat{\textbf{v}}_f(t_k)$, where
\begin{align}
	&\hat{\bm{\varphi}}_f^r(t_k)=\bigg[ \dfrac{-pB_j(p)}{A_j^2(p)}\tilde{r}(t_k), \dots, \dfrac{-p^n B_j(p)}{A_j^2(p)} \tilde{r}(t_k), \notag \\
	\label{varphir}
	& \hspace{2.3cm}\dfrac{1}{A_j(p)}\tilde{r}(t_k), \dots, \dfrac{p^m}{A_j(p)} \tilde{r}(t_k) \bigg]^\top,
\end{align}
and the contribution of the disturbance is again decomposed as $\hat{\textbf{v}}_f(t_k) = \hat{\textbf{v}}_{f1}(t_k)+ \textbf{v}_{f2}(t_k)$, with $\textbf{v}_{f2}(t_k)$ being given by \eqref{varphiv2} and
\begin{equation}
	\label{hatvarphiv1}
	\hat{\textbf{v}}_{\hspace{-0.02cm}f1}(\hspace{-0.01cm}t_k\hspace{-0.01cm})\hspace{-0.08cm}=\hspace{-0.11cm}\begin{bmatrix}
		\hspace{-0.05cm}\dfrac{-pB_{\hspace{-0.02cm}j}\hspace{-0.02cm}(\hspace{-0.01cm}p\hspace{-0.01cm})}{A_j^2(p)}\tilde{v}(t_k), & \hspace{-0.1cm}\dots, & \hspace{-0.15cm}\dfrac{-p^n \hspace{-0.04cm} B_{\hspace{-0.02cm}j}\hspace{-0.02cm}(\hspace{-0.01cm}p\hspace{-0.01cm})}{A_j^2(p)} \tilde{v}(\hspace{-0.01cm}t_k\hspace{-0.01cm}), & \hspace{-0.05cm} 0, & \hspace{-0.05cm} \dots, & \hspace{-0.03cm} 0
	\end{bmatrix}^{\hspace{-0.07cm}\top}\hspace{-0.1cm}.
\end{equation}
\subsection{Consistency of the SRIVC estimator for Setting 2}
Sufficient conditions for the generic consistency of the SRIVC estimator under Setting 2 are derived next.
\begin{thm}
	\label{theorem1}
	Consider the SRIVC estimator described by the iterations in \eqref{iterations}, filtered regressor \eqref{filteredregressor}, filtered instrument \eqref{filteredinstrument_open}, and filtered output \eqref{filteredoutput}, and suppose Assumptions \ref{assumption21} to \ref{assumption23} and \ref{assumption51} to \ref{assumption54} hold. Furthermore, assume that $v(t_k)$ is a white noise process. Then, the following statements are true:
	\begin{enumerate}
		\item The matrix $\mathbb{E}\{\hat{\bm{\varphi}}_f(t_k)\bm{\varphi}_f^\top(t_k)\}$ of the SRIVC method is generically non-singular with respect to the system and model denominator provided that the condition
		\begin{align}
			&\hspace{-0.3cm}\Big\|\mathbb{E}\Big\{\hspace{-0.05cm}\hat{\bm{\varphi}}_{\hspace{-0.02cm}f}^r\hspace{-0.03cm}(\hspace{-0.01cm}t_k\hspace{-0.01cm})\bm{\Delta}^{\hspace{-0.07cm}\top}\hspace{-0.07cm}(\hspace{-0.01cm}t_k\hspace{-0.01cm})\hspace{-0.06cm}+\hspace{-0.06cm}\hat{\mathbf{v}}_{\hspace{-0.02cm}f\hspace{-0.01cm}1}\hspace{-0.03cm}(\hspace{-0.01cm}t_k\hspace{-0.01cm})\mathbf{v}_{\hspace{-0.02cm}f}^{\hspace{-0.03cm}\top}\hspace{-0.07cm}(\hspace{-0.01cm}t_k\hspace{-0.01cm})\hspace{-0.06cm}+\hspace{-0.06cm}\mathbf{v}_{\hspace{-0.02cm}f2}\hspace{-0.03cm}(\hspace{-0.01cm}t_k\hspace{-0.01cm})\mathbf{v}_{\hspace{-0.04cm}f1}^\top\hspace{-0.03cm}(\hspace{-0.01cm}t_k\hspace{-0.01cm})\hspace{-0.04cm}\Big\}\hspace{-0.04cm}\Big\|_2\hspace{-0.05cm} \notag \\ 
			\label{conditionstatement1}
			&<\hspace{-0.05cm}\sigma_{\hspace{-0.02cm}\textnormal{min}}\hspace{-0.06cm}\left(\hspace{-0.03cm}\mathbb{E}\hspace{-0.04cm}\left\{\hspace{-0.05cm}\hat{\bm{\varphi}}_{\hspace{-0.02cm}f}^r\hspace{-0.02cm}(t_k){{}\tilde{\bm{\varphi}}_f^r}^{\hspace{-0.05cm}\top}\hspace{-0.06cm}(\hspace{-0.01cm}t_k\hspace{-0.01cm})\hspace{-0.05cm}+\hspace{-0.06cm}\mathbf{v}_{\hspace{-0.02cm}f2}\hspace{-0.03cm}(\hspace{-0.01cm}t_k\hspace{-0.01cm})\mathbf{v}_{\hspace{-0.03cm}f2}^\top\hspace{-0.03cm}(\hspace{-0.01cm}t_k\hspace{-0.01cm})\hspace{-0.04cm}\right\}\hspace{-0.02cm}\right)\hspace{-0.15cm}
		\end{align}
		holds, where $\tilde{\bm{\varphi}}_f^r(t_k)$, $\bm{\Delta}(t_k)$, $\mathbf{v}_{f1}(t_k)$, $\mathbf{v}_{f2}(t_k)$, $\hat{\bm{\varphi}}_f^r(t_k)$,and $\hat{\mathbf{v}}_{f1}(t_k)$ are defined as in \eqref{tildevarphir}, \eqref{delta}, \eqref{varphiv1}, \eqref{varphiv2}, \eqref{varphir}, and \eqref{hatvarphiv1}, respectively.
		
		\item
		If \eqref{conditionstatement1} is satisfied and the iterations of the SRIVC estimator converge for all $N$ sufficiently large to, say, $\bar{\bm{\theta}}^N$, then the true parameter $\bm{\theta}^*$ is the unique converging point of $\bar{\bm{\theta}}^N$ as $N\to \infty$.
	\end{enumerate}
\end{thm}
\textit{Proof of Statement 1}: By exploiting the decompositions of the filtered regressor and instrument vectors, the modified normal matrix 
$\mathbb{E}\{\hat{\bm{\varphi}}_f(t_k)\bm{\varphi}_f^\top(t_k)\}$ can be written as
\begin{align}
	&\mathbb{E}\hspace{-0.03cm}\{\hspace{-0.02cm}\hat{\bm{\varphi}}_{\hspace{-0.02cm}f}\hspace{-0.02cm}(\hspace{-0.02cm}t_k\hspace{-0.02cm})\bm{\varphi}_{\hspace{-0.02cm}f}^{\hspace{-0.02cm}\top}\hspace{-0.05cm}(\hspace{-0.02cm}t_k\hspace{-0.02cm})\hspace{-0.02cm}\} \hspace{-0.1cm}=\hspace{-0.08cm} \mathbb{E}\hspace{-0.06cm}\left\{\hspace{-0.06cm}\hat{\bm{\varphi}}_{\hspace{-0.02cm}f}^r\hspace{-0.02cm}(\hspace{-0.02cm}t_k\hspace{-0.02cm})\hspace{-0.02cm}\big(\hspace{-0.02cm}\tilde{\bm{\varphi}}_{\hspace{-0.03cm}f}^r\hspace{-0.03cm}(\hspace{-0.02cm}t_k\hspace{-0.02cm})\hspace{-0.08cm}+\hspace{-0.08cm}\bm{\Delta}\hspace{-0.02cm}(\hspace{-0.02cm}t_k\hspace{-0.02cm})\hspace{-0.02cm}\big)^{\hspace{-0.11cm}\top}\hspace{-0.16cm}+\hspace{-0.07cm}\hat{\mathbf{v}}_{\hspace{-0.02cm}f}\hspace{-0.03cm}(\hspace{-0.02cm}t_k\hspace{-0.02cm})\mathbf{v}_{\hspace{-0.03cm}f}^{\hspace{-0.04cm}\top}\hspace{-0.06cm}(\hspace{-0.02cm}t_k\hspace{-0.02cm})\hspace{-0.05cm}\right\} \notag \\ &\hspace{0.6cm}-\mathbb{E}\hspace{-0.03cm}\left\{\hspace{-0.04cm}\hat{\mathbf{v}}_{\hspace{-0.01cm}f}\hspace{-0.02cm}(t_k)\big(\tilde{\bm{\varphi}}_{\hspace{-0.02cm}f}^r\hspace{-0.02cm}(t_k)\hspace{-0.07cm}+\hspace{-0.06cm}\bm{\Delta}(t_k)\big)^{\hspace{-0.06cm}\top}\hspace{-0.08cm}+\hspace{-0.04cm}\hat{\bm{\varphi}}_{\hspace{-0.02cm}f}^r\hspace{-0.02cm}(t_k)\mathbf{v}_f^\top\hspace{-0.02cm}(t_k)\hspace{-0.02cm}\right\}\hspace{-0.04cm}.\notag
\end{align}
Since the external reference and disturbance are statistically independent by Assumption \ref{assumption22}, the second expectation in the equation above is equal to zero. Thus,
\begin{subequations}
	\label{secondline}
	\begin{align}
		\label{secondline1}
		\mathbb{E}\{\hat{\bm{\varphi}}_f\hspace{-0.03cm}(t_k)\bm{\varphi}_{\hspace{-0.03cm}f}^{\hspace{-0.03cm}\top}\hspace{-0.04cm}(t_k)\} \hspace{-0.09cm}&=\hspace{-0.08cm} \mathbb{E}\hspace{-0.05cm}\left\{\hspace{-0.05cm}\hat{\bm{\varphi}}_f^r(t_k){{}\tilde{\bm{\varphi}}_f^r}^{\hspace{-0.04cm}\top}\hspace{-0.03cm}(t_k)\hspace{-0.04cm}+\hspace{-0.04cm}\textbf{v}_{f2}(t_k)\textbf{v}_{f2}^\top(t_k)\right\} \\
		\label{secondline2}
		&\hspace{-2.4cm}+\mathbb{E}\Big\{\hat{\bm{\varphi}}_f^r(t_k)\bm{\Delta}^{\hspace{-0.04cm}\top}\hspace{-0.05cm}(t_k)\hspace{-0.05cm}+\hspace{-0.05cm}\hat{\mathbf{v}}_{f1}\hspace{-0.03cm}(t_k)\mathbf{v}_f^\top(t_k)\hspace{-0.05cm}+\hspace{-0.05cm}\mathbf{v}_{f2}(t_k)\mathbf{v}_{f1}^\top(t_k)\Big\}.
	\end{align}
\end{subequations}
Under condition \eqref{conditionstatement1}, the matrix in~\eqref{secondline2} is small enough (in 2-norm) to not affect the non-singularity of $\mathbb{E}\{\hat{\bm{\varphi}}_f(t_k)\bm{\varphi}_f^\top(t_k)\}$. Theorem 5.1 of \cite{dahleh2002lectures} ensures the generic non-singularity of the modified normal matrix provided that \eqref{conditionstatement1} holds and that  $\mathbb{E}\{\hat{\bm{\varphi}}_f^r(t_k){{}\tilde{\bm{\varphi}}_f^r}^\top(t_k)+\textbf{v}_{f2}(t_k)\textbf{v}_{f2}^\top(t_k)\}$ is generically non-singular.

We now prove the generic non-singularity of the matrix $\mathbb{E}\{\hat{\bm{\varphi}}_{f}^r(t_k){{}\tilde{\bm{\varphi}}_{f}^r}^{\top}(t_k)+\textbf{v}_{f2}(t_k)\textbf{v}_{f2}^\top(t_k)\}$. It is known that (cf.~\eqref{equivalent1} and \eqref{equivalent2})
\begin{align}
	\label{expressionssylvester1}
	\hat{\boldsymbol\varphi}_{f}^r(t_k) &=\mathbf{S}(-B_j,A_j)\frac{1}{A_j^2(p)} \tilde{\mathbf{r}}_{n+m}(t_k) \\
	\label{expressionssylvester2}
	\tilde{\boldsymbol\varphi}_f^r(t_k) &= \mathbf{S}(-B^*,A^*) \frac{1}{A_j(p)A^{*}(p)}\tilde{\mathbf{r}}_{n+m}(t_k),
\end{align}
where $\tilde{\mathbf{r}}_{n+m}(t_k) := [p^{n+m}, p^{n+m-1}, \dots, 1]^\top \tilde{r}(t_k),$ and the Sylvester matrices $\mathbf{S}(-B^*,A^*)$ and $\mathbf{S}(-B_j,A_j)$ are non-singular due to Lemma A3.1 of \cite{soderstrom1983instrumental}, since these polynomials are coprime by Assumptions \ref{assumption21} and~\ref{assumption52} respectively. Substituting \eqref{expressionssylvester1} and \eqref{expressionssylvester2} into the first summand in \eqref{secondline1} yields
\begin{align}
	\mathbb{E}&\left\{\hat{\bm{\varphi}}_f^r(t_k){{}\tilde{\bm{\varphi}}_f^r}^\top(t_k)+\textbf{v}_{f2}(t_k)\textbf{v}_{f2}^\top(t_k)\right\} \notag \\
	&= \mathbf{S}(-B_j,A_j) \bm{\Phi} \mathbf{S}(-B^*,A^*) + \mathbb{E}\left\{\textbf{v}_{f2}(t_k)\textbf{v}_{f2}^\top(t_k)\right\}, \notag 
\end{align}
where
\begin{equation}
	\label{phi}
	\bm{\Phi}=\mathbb{E}\bigg\{\frac{1}{A_j^2(p)}\tilde{\mathbf{r}}_{n+m}(t_k)\frac{1}{A_j(p)A^*(p)}\tilde{\mathbf{r}}_{n+m}^\top(t_k)\bigg\}.
\end{equation}
We show the generic non-singularity of $\mathbb{E}\{\hat{\bm{\varphi}}_{\hspace{-0.02cm}f}^r\hspace{-0.03cm}(\hspace{-0.01cm}t_k\hspace{-0.01cm}){{}\tilde{\bm{\varphi}}_f^r}^{\hspace{-0.04cm}\top}\hspace{-0.05cm}(\hspace{-0.01cm}t_k\hspace{-0.01cm})\hspace{-0.04cm}+\hspace{-0.04cm}\textbf{v}_{f2}(t_k)\textbf{v}_{f2}^\top(t_k)\}$ by proving that every entry of such matrix is a real analytic function in the joint variables of the model and system denominator, and that it is positive definite when evaluated at the true system parameters. The analyticity of each entry of the matrix in question follows from Lemma 9 of \cite{pan2020consistency} applied to each entry in~\eqref{phi} and $\mathbb{E}\{\textbf{v}_{f2}(t_k)\textbf{v}_{f2}^\top(t_k)\}$  and using the fact that the sum and multiplication of real analytic functions is real analytic. For the positive definiteness condition, we require that $\tilde{r}(t_k)$ is persistently exciting of order no less than $2n$ for Lemma 7 of \cite{pan2020consistency} to be applied. Since
\begin{equation}
	\frac{C_\textnormal{d}(q)}{1+G_\textnormal{d}^*(q)C_\textnormal{d}(q)} = \frac{A_\textnormal{d}^*(q)F_\textnormal{d}(q)}{A_\textnormal{d}^*(q)L_\textnormal{d}(q)+B_\textnormal{d}^*(q)F_\textnormal{d}(q)}, \notag 
\end{equation}
we find that at least $n$ zeros of this transfer function are minimum-phase (namely, the zeros of $A_\textnormal{d}^*(q)$). Thus, Assumption \ref{assumption53} ensures that the spectrum of $\tilde{r}(t_k)$ has at least $2n$ distinct frequency lines, and therefore Lemma 7 of \cite{pan2020consistency} ensures that $\bm{\Phi}$ is positive definite when evaluated at the true parameters. Since  $\mathbb{E}\{\textbf{v}_{f2}(t_k)\textbf{v}_{f2}^\top(t_k)\}$ is positive semi-definite by construction, we have that $\mathbb{E}\{\hat{\bm{\varphi}}_f^r(t_k){{}\tilde{\bm{\varphi}}_f^r}^\top(t_k)+\textbf{v}_{f2}(t_k)\textbf{v}_{f2}^\top(t_k)\}$ is positive definite when evaluated at the true parameters. With this, by Lemma \ref{lemmageneric} we obtain that the matrix $\mathbb{E}\{\hat{\bm{\varphi}}_f^r(t_k){{}\tilde{\bm{\varphi}}_f^r}^\top(t_k)+\textbf{v}_{f2}(t_k)\textbf{v}_{f2}^\top(t_k)\}$ is generically non-singular. This concludes the proof of Statement 1. \hfill $\square$\\

\textit{Proof of Statement 2}: The ergodicity results in \cite[Lemma 3.1]{soderstrom1975ergodicity} and \cite[Lemma A4.3]{soderstrom1983instrumental} permit us to write~\eqref{iterations}, at the converging point and as $N$ tends to infinity, as
\begin{equation}
	\bar{\bm{\theta}} = \mathbb{E}\left\{\hat{\bm{\varphi}}_f(t_k,\bar{\bm{\theta}}) \bm{\varphi}_f^\top(t_k,\bar{\bm{\theta}})\right\}^{-1} \mathbb{E}\left\{\hat{\bm{\varphi}}_f(t_k,\bar{\bm{\theta}}) y_f(t_k,\bar{\bm{\theta}})\right\}, \notag
\end{equation}
where $\bar{A}(p)$ and $\bar{B}(p)$ are the corresponding denominator and numerator polynomials of the converging point as $N\to \infty$. These polynomials are coprime by Assumption \ref{assumption52}. Since the matrix inverse in the equation above is non-singular by Statement 1, the following must hold:
\begin{equation}
	\label{condition1}
	\mathbb{E}\left\{\hat{\bm{\varphi}}_f(t_k,\bar{\bm{\theta}}) \big(y_f(t_k,\bar{\bm{\theta}})- \bm{\varphi}_f^\top(t_k,\bar{\bm{\theta}})\bar{\bm{\theta}}\big)\right\} = \mathbf{0}.
\end{equation}
From \eqref{filteredregressor} and \eqref{filteredoutput}, after some computations we find that
\begin{align}
	y_f(t_k,\bar{\bm{\theta}})- \bm{\varphi}_f^\top(t_k,\bar{\bm{\theta}})\bar{\bm{\theta}} = \big(G^*(p)-\bar{G}(p)\big)u(t_k)+v(t_k). \notag
\end{align}
By Lemma \ref{lemmacor} (see Appendix), $\mathbb{E}\{\hat{\bm{\varphi}}_f(t_k,\bar{\bm{\theta}})v(t_k)\}\hspace{-0.08cm}=\hspace{-0.06cm}\mathbf{0}$ and therefore \eqref{condition1} is equivalent to
\begin{equation}
	\mathbb{E}\left\{\hat{\bm{\varphi}}_f(t_k,\bar{\bm{\theta}}) [G^*(p)-\bar{G}(p)]\big(\tilde{r}(t_k)-\tilde{v}(t_k)\big)\right\} = \mathbf{0}, \notag
\end{equation}
where the notation in \eqref{tilderv} has been used. After standard computations, we see that
\begin{align}
	[G^*(p)&-\bar{G}(p)]\big(\tilde{r}(t_k)-\tilde{v}(t_k)\big) \notag \\
	\label{exploiting1}
	&=\frac{1}{A^*(p)\bar{A}(p)} \big(\tilde{\mathbf{r}}_{n+m}^\top(t_k)-\tilde{\mathbf{v}}_{n+m}^\top(t_k)\big)\mathbf{h},
\end{align}
where $\tilde{\mathbf{v}}_{n+m}(t_k)$ has the same form as $\tilde{\mathbf{r}}_{n+m}(t_k)$ but $\tilde{r}(t_k)$ is replaced by $\tilde{v}(t_k)$. The vector $\mathbf{h}$ is formed by the coefficients of the polynomial $\bar{A}(p)B^*(p)-\bar{B}(p)A^*(p)$. On the other hand, we can express the filtered instrument as
\begin{equation}
	\label{exploiting2}
	\hat{\bm{\varphi}}_f(t_k,\bar{\bm{\theta}}) = \mathbf{S}(-\bar{B},\bar{A}) \frac{1}{\bar{A}^2(p)}\big(\tilde{\mathbf{r}}_{n+m}(t_k)-\tilde{\mathbf{v}}_{n+m}(t_k)\big).
\end{equation}
Since $\mathbf{S}(-\bar{B},\bar{A})$ is non-singular and the external reference signal $r(t_k)$ and disturbance $v(t_k)$ are independent by Assumption \ref{assumption22}, we can use \eqref{exploiting1} and \eqref{exploiting2} to write condition \eqref{condition1} as $\bar{\bm{\Phi}} \mathbf{h}=\mathbf{0}$, with
\begin{align}
	\bar{\bm{\Phi}}&=\mathbb{E}\left\{\frac{1}{\bar{A}^2(p)}\tilde{\mathbf{r}}_{n+m}(t_k)\frac{1}{\bar{A}(p)A^*(p)}\tilde{\mathbf{r}}_{n+m}^\top(t_k)\right\} \notag \\
	\label{phibar}
	&\hspace{0.3cm}+ \hspace{-0.04cm}\mathbb{E}\hspace{-0.04cm}\left\{\hspace{-0.04cm}\frac{1}{\bar{A}^2(p)}\tilde{\mathbf{v}}_{n+m}(t_k)\frac{1}{\bar{A}(p)A^*(p)}\tilde{\mathbf{v}}_{n+m}^\top(t_k)\hspace{-0.04cm}\right\}.
\end{align}
The generic non-singularity of $\bar{\bm{\Phi}}$ follows from the same procedure as in the proof of Statement 1. Thus, we must have $\mathbf{h}=\mathbf{0}$, implying $\bar{G}(p)=G^*(p)$, i.e., $\bm{\theta}^*$ is the unique limiting point. \hfill $\square$

\begin{rem}
	The condition in \eqref{conditionstatement1} is sufficient but not necessary for the generic non-singularity of the matrix $\mathbb{E}\{\hat{\bm{\varphi}}_f(t_k)\bm{\varphi}_f^\top(t_k)\}$. This requirement can always be satisfied as long as the signal-to-noise ratio (SNR) between the external reference and the disturbance is high enough, and the interpolation error $\mathbf{\Delta}(t_k)$ is not significant. 
\end{rem}

Corollary \ref{corollary1} shows that generic consistency is lost if there is a direct feedthrough of the loop transfer, or if the noise is not white. We assume that the modified normal matrix is non-singular; this fact can be proven generically using the same tools developed in this paper, provided that a condition similar to \eqref{conditioninv} or \eqref{conditionstatement1} is satisfied.
\begin{cor}
	\label{corollary1}
	Assume that the SRIVC iterations converge for all $N$ sufficiently large and that $\mathbb{E}\{\hat{\bm{\varphi}}_f(t_k)\bm{\varphi}_f^\top(t_k)\}$ is non-singular. Under Assumptions \ref{assumption21} to \ref{assumption23} and \ref{assumption52} to \ref{assumption54}, if $v(t_k)$ is not white noise, or if Assumption~\ref{assumption51} is not satisfied, then generically $\bar{G}(p)\neq G^*(p)$ as $N\to\infty$.
\end{cor}	
\begin{pf*}{Proof.}
	Following the same steps as in the proof of Statement 2 of Theorem \ref{theorem1}, we find that the limiting point of the SRIVC estimator must satisfy, as $N$ tends to infinity,
	\begin{equation}
		\mathbb{E}\hspace{-0.03cm}\left\{\hspace{-0.03cm}\hat{\bm{\varphi}}_f(t_k,\hspace{-0.02cm}\bar{\bm{\theta}}) [G^*\hspace{-0.03cm}(p)\hspace{-0.07cm}-\hspace{-0.04cm}\bar{G}(p)]u(t_k)\hspace{-0.02cm}\right\} \hspace{-0.06cm}=\hspace{-0.06cm} -\mathbb{E}\left\{\hspace{-0.03cm}\hat{\bm{\varphi}}_f(t_k,\bar{\bm{\theta}}) v(t_k)\right\}\hspace{-0.04cm}. \notag
	\end{equation}
	By exploiting \eqref{exploiting1}, \eqref{exploiting2} and \eqref{lefthandside}, we can rewrite the equation above as
	\begin{equation}
		\bar{\bm{\Phi}} \mathbf{h} \hspace{-0.06cm}=\hspace{-0.06cm} \mathbb{E}\hspace{-0.05cm}\left\{\hspace{-0.06cm}\left[\frac{\bar{\mathbf{B}}_{n+m}^{\textnormal{d}}(q)C_\textnormal{d}(q)}{\bar{A}_\textnormal{d}^2(q)\big(1+G_\textnormal{d}^*(q)C_\textnormal{d}(q)\big)} v(t_k)\hspace{-0.02cm}\right] \hspace{-0.05cm} v(t_k)\hspace{-0.05cm}\right\}, \notag
	\end{equation}
	where $\bar{\mathbf{B}}_{n+m}^{\textnormal{d}}(q)$ is defined in \eqref{Bbar}. Since $\bar{\bm{\Phi}}$ is generically non-singular and the right-hand side is different from zero in general, we have $\mathbf{h}\neq \mathbf{0}$, i.e., $\bar{G}(p)\neq G^*(p)$ as $N$ tends to infinity. \hspace*{\fill} \qed
\end{pf*}

\begin{rem}
	A well-known fact in discrete-time system identification is that the model is biased towards the negative inverse of the controller in some closed-loop settings \cite{goodwin2002bias}. Since this aspect is intrinsic to the feedback loop configuration, it is not surprising to find similar relationships in hybrid settings with discrete-time controllers and continuous-time systems. In fact, again assuming the non-singularity of $\mathbb{E}\{\hat{\bm{\varphi}}_f(t_k) \bm{\varphi}_f^\top(t_k)\}$, the converging point must satisfy, as $N$ tends to infinity,
	\begin{align}
		&\mathbb{E}\hspace{-0.03cm}\left\{\hspace{-0.04cm}\frac{1}{\bar{A}^2(p)} \tilde{\mathbf{r}}_{n+m}\hspace{-0.02cm}(t_k)G_{\hspace{-0.04cm}\Delta}\hspace{-0.04cm}(\hspace{-0.01cm}p\hspace{-0.01cm})\tilde{r}(t_k)\hspace{-0.04cm}\right\} \notag \\
		&\hspace{-0.1cm}+\hspace{-0.06cm}\mathbb{E}\hspace{-0.06cm}\left\{\hspace{-0.1cm}\frac{1}{\bar{A}^2\hspace{-0.03cm}(\hspace{-0.01cm}p\hspace{-0.01cm})} \hspace{-0.03cm}\tilde{\mathbf{v}}_{n\hspace{-0.02cm}+\hspace{-0.02cm}m}\hspace{-0.04cm}(\hspace{-0.01cm}t_k\hspace{-0.01cm})\hspace{-0.02cm}G_{\hspace{-0.04cm}\Delta}\hspace{-0.04cm}(\hspace{-0.01cm}p\hspace{-0.01cm})\tilde{v}(\hspace{-0.01cm}t_k\hspace{-0.01cm})\hspace{-0.06cm}\right\}\hspace{-0.1cm}=\hspace{-0.07cm}\mathbb{E}\hspace{-0.06cm}\left\{\hspace{-0.09cm}\frac{1}{\bar{A}^2\hspace{-0.03cm}(\hspace{-0.01cm}p\hspace{-0.01cm})}\hspace{-0.04cm} \tilde{\mathbf{v}}_{n\hspace{-0.02cm}+\hspace{-0.02cm}m}\hspace{-0.04cm}(\hspace{-0.01cm}t_k\hspace{-0.01cm})\tilde{v}(\hspace{-0.01cm}t_k\hspace{-0.01cm})\hspace{-0.07cm}\right\}\hspace{-0.06cm}, \notag
	\end{align}
	where $G_\Delta(p)=G^*(p)-\bar{G}(p)$. After some computations, this equation can be rewritten as
	\begin{align}
		&\mathbf{0} = \mathbb{E}\left\{\frac{1}{\bar{A}^2(p)} \tilde{\mathbf{r}}_{n+m}(t_k)G_\Delta(p)\tilde{r}(t_k)\right\} \notag \\
		\label{tobeexpected}
		&\hspace{-0.1cm}+ \mathbb{E}\left\{\frac{1}{\bar{A}^2(p)} \tilde{\mathbf{v}}_{n+m}(t_k)[-C_\textnormal{d}^{-1}(q)-\bar{G}_{\textnormal{d}}(q)]\tilde{v}(t_k)\right\}.
	\end{align}
	Thus, we find that the converging transfer function of the SRIVC estimator will tend towards $G^*(p)$ for high SNR. Indeed, for that case, the first expectation in~\eqref{tobeexpected} dictates the value of $\bar{G}(p)$ and by \eqref{exploiting1} we reach $\bar{\bm{\Phi}}_1 \mathbf{h}\approx\mathbf{0}$, where $\bar{\bm{\Phi}}_1$ is the first expectation in \eqref{phibar}. Hence, we obtain $\bar{G}(p)\approx G^*(p)$. On the contrary, for low SNR the second expectation in \eqref{tobeexpected} impacts the value of $\bar{G}(p)$ the most, and the zero-order hold equivalent of $\bar{G}(p)$ will tend to $-1/C_\textnormal{d}(q)$ by a similar reasoning as before. 
\end{rem}

\subsection{The CLSRIVC estimator for Setting 2 and its consistency analysis}
\label{clsrivcsetup2}
The CLSRIVC estimator, as introduced in \cite{garnier2008book}, assumes that a \textit{continuous-time} controller is known in advance for the implementation of the algorithm. Therefore, it is only suitable for Setting 1. In this work we introduce and analyze an extension of this estimator which is valid for Setting 2 that has a filtered instrument given by
\begin{align}
	&\hat{\bm{\varphi}}_{f}(t_k,\bm{\theta}_j) = \bigg[\frac{-p B_j(p)}{A_j^2(p)} ,\hspace{0.05cm} \dots, \hspace{0.05cm}\frac{-p^n B_j(p)}{A_j^2(p)}, \notag \\
	\label{filteredinstrument_closed2}
	& \hspace{2cm}\frac{1}{A_j(p)}, \hspace{0.05cm}\dots, \hspace{0.05cm}\frac{p^m}{A_j(p)}\bigg]^\top S_{uo,j}(q) r(t_k),
\end{align} 
where $S_{uo,j}(q)=C_\textnormal{d}(q)/[1+G_{\textnormal{d},j}(q)C_\textnormal{d}(q)]$. The other signals of interest at each iteration ($\bm{\varphi}_{f}(t_k,\bm{\theta}_j)$ and $y_{f}(t_k,\bm{\theta}_j)$) have the same form as the standard SRIVC estimator, that is, they are given by \eqref{filteredregressor} and \eqref{filteredoutput}, and the iterations are computed by \eqref{iterations}. We prove the generic consistency of this variant of the CLSRIVC estimator for Setting 2 in Theorem \ref{theorem2}.
\begin{thm}\hspace{-0.2cm}\textbf{.}
	\label{theorem2}
	Consider the CLSRIVC estimator described by the iterations in~\eqref{iterations} with filtered instrument vector \eqref{filteredinstrument_closed2}, and suppose Assumptions \ref{assumption21} to \ref{assumption23} and \ref{assumption51} to \ref{assumption54} hold. Then, the following statements are true:
	\begin{enumerate}
		\item The matrix $\mathbb{E}\{\hat{\bm{\varphi}}_f(t_k)\bm{\varphi}_f^\top(t_k)\}$ is generically non-singular with respect to the system and model denominator provided that the condition
		\begin{equation}
			\label{conditionstatement1_clsrivc}
			\big\|\mathbb{E}\hspace{-0.04cm}\left\{\hspace{-0.02cm}\hat{\boldsymbol{\varphi}}_f(t_k)\boldsymbol{\Delta}^{\hspace{-0.04cm}\top}\hspace{-0.04cm}(t_k)\right\}\hspace{-0.08cm}\big\|_2\hspace{-0.06cm}<\hspace{-0.03cm}\sigma_{\textnormal{min}}\hspace{-0.05cm}\left(\mathbb{E}\big\{\hat{\boldsymbol{\varphi}}_f(t_k){{}{\tilde{\boldsymbol{\varphi}}_f^r}}^\top(t_k)\big\}\right)
		\end{equation}
		holds, where $\tilde{\bm{\varphi}}_{f}^r(t_k), \bm{\Delta}(t_k)$ and $\hat{\bm{\varphi}}_{f}(t_k)$ are defined as in \eqref{tildevarphir}, \eqref{delta}, and~\eqref{filteredinstrument_closed2}, respectively.
		\item
		If \eqref{conditionstatement1_clsrivc} is satisfied and the iterations of the CLSRIVC estimator converge for all $N$ sufficiently large to, say, $\bar{\bm{\theta}}^N$, then the true parameter $\bm{\theta}^*$ is the unique converging point of $\bar{\bm{\theta}}^N$ as $N\to \infty$.
	\end{enumerate}
\end{thm}	

\textit{Proof of Statement 1}: Statement 1 follows from a similar reasoning to the proof of Lemma \ref{lemmainv} the application of Theorem 5.1 of \cite{dahleh2002lectures}. This time,
\begin{equation}
	\boldsymbol\Phi \hspace{-0.1cm}=\hspace{-0.1cm} \mathbb{E}\hspace{-0.08cm}\left\{\hspace{-0.1cm}\bigg[\hspace{-0.04cm}S_{uo,\hspace{-0.02cm}j}\hspace{-0.03cm}(\hspace{-0.02cm}q\hspace{-0.02cm})\hspace{-0.02cm}\frac{1}{A_{\hspace{-0.02cm}j}^{\hspace{-0.02cm}2}\hspace{-0.04cm}(\hspace{-0.02cm}p\hspace{-0.02cm})} {\mathbf{r}_{\hspace{-0.03cm}n\hspace{-0.02cm}+\hspace{-0.02cm}m}}\hspace{-0.03cm}(\hspace{-0.02cm}t_{\hspace{-0.01cm}k}\hspace{-0.02cm})\hspace{-0.03cm}\bigg] \hspace{-0.12cm}\left[\hspace{-0.05cm}S_{\hspace{-0.03cm}uo}^*\hspace{-0.03cm}(\hspace{-0.01cm}q\hspace{-0.01cm})\hspace{-0.02cm}\frac{1}{A_{\hspace{-0.02cm}j}\hspace{-0.04cm}(\hspace{-0.02cm}p\hspace{-0.02cm})\hspace{-0.02cm}A^{\hspace{-0.02cm}*}\hspace{-0.04cm}(\hspace{-0.02cm}p\hspace{-0.02cm})} \hspace{-0.02cm}\mathbf{r}_{\hspace{-0.04cm}n\hspace{-0.02cm}+\hspace{-0.02cm}m}^\top\hspace{-0.03cm}(\hspace{-0.02cm}t_{\hspace{-0.01cm}k}\hspace{-0.02cm})\hspace{-0.02cm}\right]\hspace{-0.1cm}\right\}\hspace{-0.08cm}. \notag
\end{equation}
Note that the continuous-time transfer functions involved in the construction of the elements of $\boldsymbol\Phi$ are well defined when evaluating at the true parameters due to Assumption \ref{assumption23}. Furthermore, the matrix $\bm{\Phi}$ above is equivalent to~\eqref{phi} when they are evaluated at the true parameters. Thus, its positive-definiteness follows from the persistence of excitation condition in Assumption \ref{assumption53}, and the analyticity of the entries of $\bm{\Phi}$ again follow from Lemma 9 of \cite{pan2020consistency}. Thus, we conclude that the modified normal matrix $\mathbb{E}\{\hat{\boldsymbol\varphi}_f(t_k)\boldsymbol\varphi^\top_f(t_k)\}$ is generically non-singular by Lemma \ref{lemmageneric}. \hfill $\square$

\textit{Proof of Statement 2}: The first lines of the proof of Statement 2 are the same as in the proof of Theorem~\ref{thmclsrivc_consistency} until \eqref{untilthispoint}. Since the input $u(t)$ is perfectly reconstructed from a ZOH device, we have $\{G^*(p)u(t)\}_{t=t_k}=G^*(p)u(t_k)$ and thus $\mathbb{E}\big\{\hat{\boldsymbol\varphi}_f(t_k,\bar{\boldsymbol\theta}) \left([G^*(p)-\bar{G}(p)]\tilde{r}(t_k)\right)\hspace{-0.08cm}\big\} = \mathbf{0}.$ Exploiting the expressions in \eqref{exploiting1} and \eqref{exploiting2}, together with the non-singularity of $\mathbf{S}(-\bar{B},\bar{A})$, yields
\begin{equation}
	\mathbb{E}\bigg\{\frac{1}{\bar{A}^2(p)}\tilde{\mathbf{r}}_{n+m}(t_k)\frac{1}{\bar{A}(p)A^*(p)}\tilde{\mathbf{r}}_{n+m}^\top(t_k)\bigg\} \mathbf{h}=\mathbf{0}. \notag
\end{equation}
The matrix above is known to be generically non-singular by leveraging the same ideas as in Statement 1. Therefore we must have $\mathbf{h}=\mathbf{0}$, implying $\bar{A}(p)B^*(p)-\bar{B}(p)A^*(p)= 0$ or equivalently $\bar{G}(p)=G^*(p)$, i.e., $\bm{\theta}^*$ is the unique limiting point. \hfill $\square$

In summary, we have found that the SRIVC and CLSRIVC estimators are generically consistent for Setting 2. Contrary to Setting 1, the estimators do not require oversampling to alleviate bias issues since the intersample behavior of the system input is known. The SRIVC estimator requires stricter conditions in the model and noise for consistency. When these are not satisfied, the discrete-time equivalent of the estimate is biased towards the negative inverse of the controller.

\section{Simulations}
\label{simulations}
We now corroborate the theoretical results via several Monte Carlo simulations. This section is divided in two parts: the consistency tests (that is, the verification of Theorems \ref{thmsrivc_consistency}, \ref{thmclsrivc_consistency}, \ref{theorem1} and \ref{theorem2}), and an example of the bias behavior of the SRIVC estimator for Setting 2.

\subsection{Consistency tests}
We consider the continuous-time system
\begin{equation}
	G^*(p) = \frac{-0.25p+0.5}{0.5p^2+0.707p+1}. \notag 
\end{equation} 
The sampling period is chosen as $h=0.1$[s] for all the tests, and the reference signal is white noise of variance 1 interpolated with a ZOH device. The output noise is also white, of variance $0.01$ for both settings. This variance leads to a output SNR of approximately $6.7$ [dB] for Setting 1, and approximately $9.4$ [dB] for Setting 2. The consistency of the SRIVC and CLSRIVC estimators is investigated for an increasing sample size, and they are compared with their extended versions SRIVC-os and CLSRIVC-os \cite{gonzalez2021srivc} that incorporate the exact intersample behavior of the input by oversampling it. The sample size $N$ is adjusted from $200$ to $200000$ in a logarithmic scale, where a total of 40 different sample sizes are considered. Three hundred Monte Carlo runs are performed for each sample size, and the sample mean for each parameter is recorded. The estimators are initialized with the LSSVF estimator \cite{garnier2008book}. The maximum number of iterations of each method is set to 200, and the algorithms terminate if $\|\bm{\theta}_{j+1}-\bm{\theta}_j\|_2/\|\bm{\theta}_j\|_2<10^{-7}.$ The extended estimators consider input data that is sampled $100$ times faster than the output. The following controllers are used for Settings 1 and 2 respectively:
\begin{equation}
	C(p)\hspace{-0.03cm}=\hspace{-0.04cm} 1.896\cdot 10^{-4} + \frac{0.7278}{p}, \hspace{0.15cm} C_\textnormal{d}(q) \hspace{-0.03cm}= \hspace{-0.03cm}\frac{0.416(q\hspace{-0.03cm}-\hspace{-0.03cm}0.7452)}{q-1}. \notag
\end{equation}

\begin{figure}
	\centering{
		\includegraphics[width=0.47\textwidth]{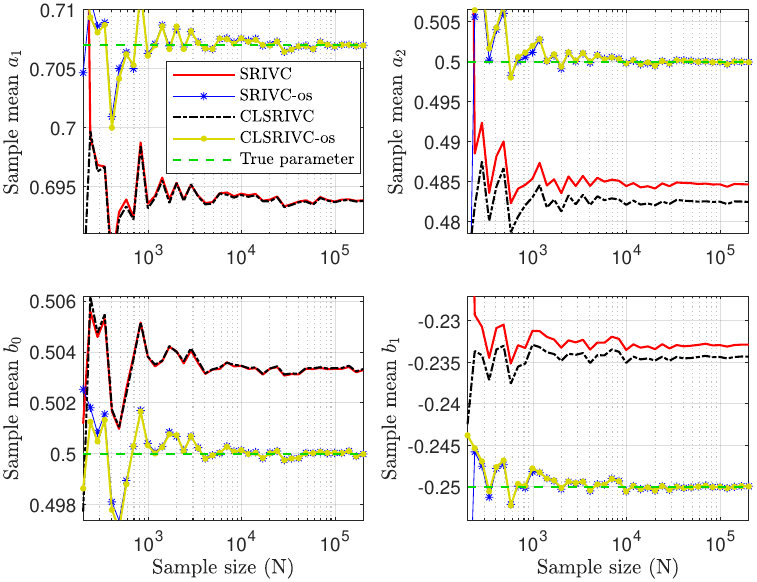}
		\caption{Mean of the estimated parameters for an increasing sample size, Setting 1.}
		\label{fig2}}
\end{figure} 
Figure \ref{fig2} shows that the SRIVC and CLSRIVC estimators are not consistent under Setting~1, as the sample means of the estimated parameters do not approach the true parameter values when the sample size increases. This observation is aligned with Theorems \ref{thmsrivc_consistency} and \ref{thmclsrivc_consistency}. In contrast, the SRIVC-os and CLSRIVC-os estimators do not exhibit noticeable bias for large sample size. 

Note that the bias of SRIVC and CLSRIVC can also be reduced by jointly decreasing the sampling period of the input and output signals. This fact can be verified theoretically for the CLSRIVC estimator from observing that the term $\varepsilon_r(t_k,\bar{\bm{\theta}})$ in~\eqref{varepsilonreference} (and therefore $\bar{\bm{\Psi}}$ in~\eqref{hcomputation}) is less significant for smaller sampling periods. In fact, if $N=200000$ and $h=0.02$[s] instead of $0.1$[s], we obtain the sample means presented in Table~\ref{table1}. The bias has been reduced as anticipated, but we should not expect it to decay to zero. Identification with small sampling periods might explain why some authors have considered the CLSRIVC estimator to be asymptotically unbiased when analyzing its performance through simulations \cite{gilson2008instrumental}. We clarify that this property does not hold from a theoretical standpoint for any non-zero sampling period.

\begin{table}
	\caption{Sample mean of the model parameters obtained with 300 Monte Carlo runs with $N=200000$ and $h=0.02$[s], Setting~1. The output SNR is approximately $2.4$ [dB].}
	\centering
	\scriptsize 
	\label{table1}
	\begin{tabular}{|c||c|c|c|c|}
		\hline
		Method  & $a_1$ ($0.707$)  & $a_2$ ($0.5$) & $b_0$ ($0.5$)  & $b_1$ ($-0.25$)   \\ \hline
		\hline
		SRIVC &  $0.7044$  & $0.4966$ & $0.5008$  & $-0.2469$ \\ \hline
		CLSRIVC  & $0.7044$  & $0.4965$ & $0.5008$  & $-0.2469$ \\
		\hline 
	\end{tabular}
\end{table}

\begin{figure}
	\centering{
		\includegraphics[width=0.47\textwidth]{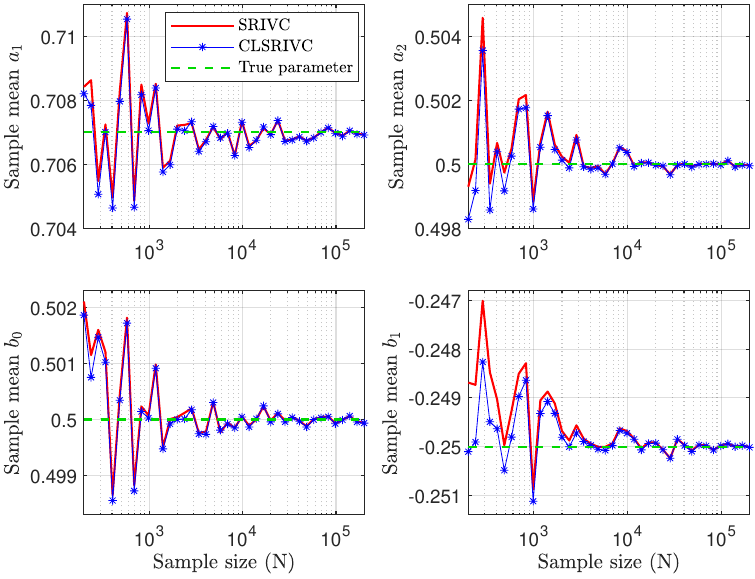}
		\caption{Mean of the estimated parameters for an increasing sample size, Setting 2.}
		\label{fig4}}
\end{figure}

\begin{figure}
	\centering{
		\includegraphics[width=0.47\textwidth]{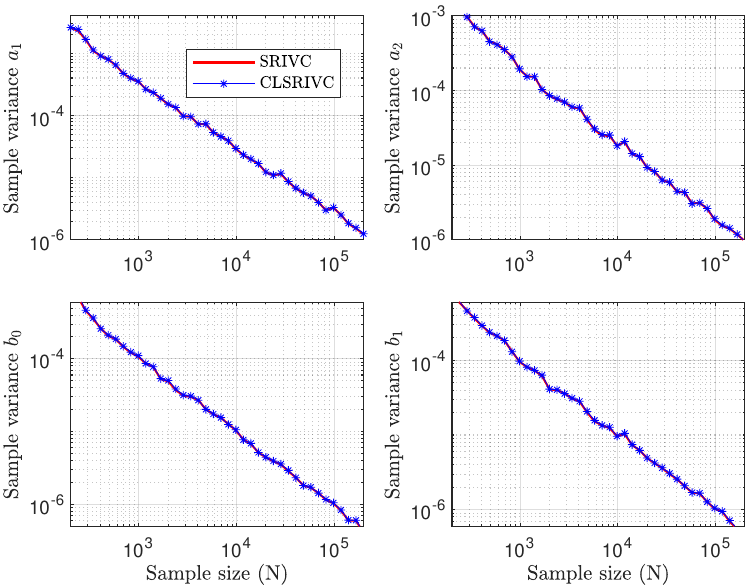}
		\caption{Variance of the estimated parameters for an increasing sample size, Setting~2.}
		\label{fig5}}
\end{figure}

In another simulation study, we test the consistency of the methods under Setting~2. The intersample behavior of $u(t_k)$ is now correctly specified in the regressor vector of the algorithms, thus yielding generically consistent estimators according to Theorems \ref{theorem1} and \ref{theorem2}. Indeed, Figures \ref{fig4} and \ref{fig5} show that the sample mean of each parameter estimate converges to the true value for an increasing sample size, and that the variance of those estimates is decreasing. This provides empirical evidence to the consistency results of Section \ref{setup2}. Note that although the SRIVC estimator is consistent in this simulation test, its consistency in closed-loop is limited to cases where the output noise is white and there is no direct feedthrough in the loop. In contrast, the CLSRIVC estimator can handle colored noise and biproper transfer functions without a loss in generic consistency.


\subsection{Bias behavior of the SRIVC estimator in Setting 2}

\begin{figure}
	\centering{
		\includegraphics[width=0.45\textwidth]{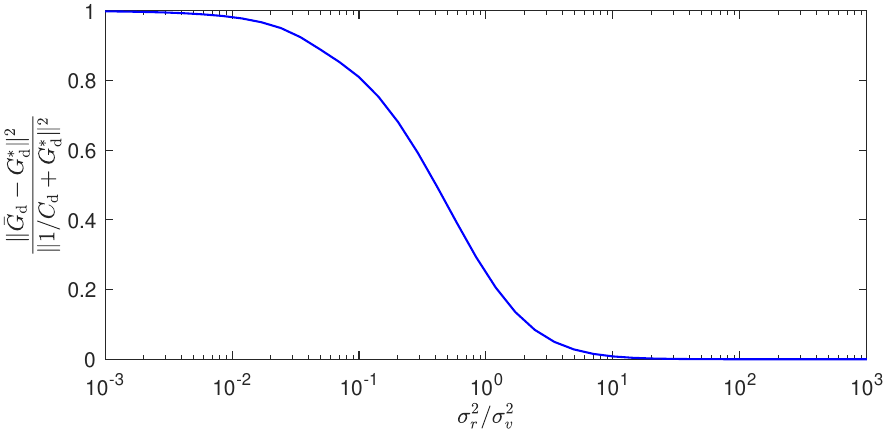}
		\caption{Normalized bias of the SRIVC estimator versus SNR, Setting 2.}
		\label{fig6}}
\end{figure}

To conclude this section, we study the bias of the SRIVC estimator when $G_\textnormal{d}^*(q) C_\textnormal{d}(q)$ is biproper. Based on \eqref{tobeexpected} we expect a bias of the ZOH equivalent of the model estimates toward $-1/C_\textnormal{d}(q)$. We verify this property for the following system and controller:
\begin{equation}
	G^*(p) = \frac{-0.3p+1}{p+1}, \quad C_\textnormal{d}(q) = \frac{2.15(q-0.9949)}{q-1}, \notag
\end{equation}
with $h=0.1$[s] and $N=100000$. Forty values that are logarithmically-spaced and ranging from $10^{-3}$ to $10^3$ are considered for the SNR, which is defined here as the quotient of the reference variance $\sigma_r^2$ and the disturbance variance $\sigma_v^2$. An average model is computed among 300 Monte Carlo runs for each SNR value, and Figure \ref{fig6} reveals the normalized bias behavior of the discrete-time equivalent with respect to the SNR in the feedback loop. As expected, the ZOH equivalent of the mean model from the SRIVC method approaches $-1/C_\textnormal{d}(q)$ for low SNR and there is a transition towards unbiasedness for high SNR. This curve, now obtained in a hybrid closed-loop scenario, is aligned with what is known for indirect non-parametric discrete-time estimators in closed-loop (see, e.g., Fig. 4 of \cite{welsh2002finite}).

\section{Conclusions}
\label{conclusions}
This paper has focused on the generic consistency properties of the SRIVC and CLSRIVC estimators for two well-established closed-loop settings. The results show that these estimators are generically consistent when there is a discrete-time controller in the loop, and this consistency is lost if a continuous-time controller is implemented. The bias for the fully continuous-time case may be partially overcome by oversampling methods that compute a more adequate regressor vector. The consistency of the CLSRIVC estimator has been shown to extend to more general scenarios than that of the SRIVC estimator, such as when there is a direct feedthrough in the loop transfer or when the output disturbance is not white. The bias of the SRIVC estimator has also been characterized in the presence of a discrete-time control feedback for such scenarios. Monte Carlo simulation tests have verified the theoretical findings.

\section*{Acknowledgments}
This work was supported by the Swedish Research Council under contract number 2016-06079 (NewLEADS), and by the Digital Futures project EXTREMUM.      

\section*{Appendix}
\begin{lem}
	\label{lemmacor}
	Assume that $v(t_k)$ is a white noise process and that Assumptions \ref{assumption22} and \ref{assumption51} hold. Then, $\mathbb{E}\{\hat{\bm{\varphi}}_f(t_k,\bar{\bm{\theta}})v(t_k)\}=\mathbf{0}$.
\end{lem}
\begin{pf*}{Proof.}
	We can write the filtered instrument as
	\begin{equation}
		\hat{\bm{\varphi}}_f(t_k,\bar{\bm{\theta}}) = \mathbf{S}(-\bar{B},\bar{A}) \frac{1}{\bar{A}_\textnormal{d}^2(q)}\mathbf{u}_{n+m}^\textnormal{d}(t_k), \notag
	\end{equation}
	where $\bar{A}_\textnormal{d}(q)$ is the denominator of the ZOH equivalent of $\bar{B}(p)/\bar{A}(p)$, and the vector $\mathbf{u}_{n+m}^\textnormal{d}(t_k)$ is given by
	\begin{equation}
		\label{Bbar}
		\mathbf{u}_{n\hspace{-0.01cm}+\hspace{-0.01cm}m}^\textnormal{d}\hspace{-0.04cm}(t_k) \hspace{-0.08cm}=\hspace{-0.1cm} 
		\underbrace{\begin{bmatrix}
				\hspace{-0.02cm}\bar{B}_\textnormal{d}^{n\hspace{-0.01cm}+\hspace{-0.01cm}m}\hspace{-0.03cm}(q), & \hspace{-0.1cm} \bar{B}_\textnormal{d}^{n\hspace{-0.01cm}+\hspace{-0.01cm}m\hspace{-0.01cm}-\hspace{-0.01cm}1}\hspace{-0.02cm}(q), & \hspace{-0.1cm} \dots, & \hspace{-0.1cm} \bar{B}_\textnormal{d}^0\hspace{-0.02cm}(q)
			\end{bmatrix}^{\hspace{-0.07cm}\top}}_{=:\bar{\mathbf{B}}_{n+m}^\textnormal{d}(q)} \hspace{-0.12cm} u(t_k), \hspace{-0.2cm}
	\end{equation}
	with $\bar{B}^{i}_\textnormal{d}(q)/\bar{A}_\textnormal{d}^2(q)$ being the ZOH equivalent of $p^{i}/\bar{A}^2(p)$, $i=0,1,\dots,n+m$. Thus, 
	\begin{equation}
		\hat{\bm{\varphi}}_f(t_k,\bar{\bm{\theta}}) = \mathbf{S}(-\bar{B},\bar{A}) \frac{\bar{\mathbf{B}}_{n+m}^\textnormal{d}(q)}{\bar{A}_\textnormal{d}^2(q)}\big(\tilde{r}(t_k) - \tilde{v}(t_k)\big). \notag
	\end{equation}
	By exploiting Assumption \ref{assumption22}, we can write the expected value $\mathbb{E}\{\hat{\bm{\varphi}}_f(t_k,\bar{\bm{\theta}})v(t_k)\}$ as
	\begin{align}
		\label{lefthandside}
		\mathbb{E}&\left\{\hat{\bm{\varphi}}_f(t_k,\bar{\bm{\theta}})v(t_k)\right\}  \\
		\notag =& \mathbf{S}(\hspace{-0.02cm}-\hspace{-0.02cm}\bar{B},\hspace{-0.02cm}\bar{A})\mathbb{E}\hspace{-0.05cm}\left\{\hspace{-0.08cm}\left[\hspace{-0.04cm}\frac{-\bar{\mathbf{B}}_{n+m}^{\textnormal{d}}(q)C_{\textnormal{d}}(q)}{\bar{A}_{\textnormal{d}}^2\hspace{-0.02cm}(q)\big(1\hspace{-0.06cm}+\hspace{-0.06cm}G_{\textnormal{d}}^*\hspace{-0.02cm}(q)C_{\textnormal{d}}(q)\big)} v(t_k)\hspace{-0.02cm}\right]\hspace{-0.08cm} v(t_k)\hspace{-0.07cm}\right\}\hspace{-0.06cm}.
	\end{align}
	Since $G_\textnormal{d}^*(q)C_\textnormal{d}(q)$ is strictly proper by Assumption \ref{assumption51}, we have $n>m$ (which means that $\bar{\mathbf{B}}_{n+m}^{\textnormal{d}}(q)/\bar{A}_\textnormal{d}^2(q)$ is formed solely by strictly proper transfer functions), or $C_\textnormal{d}(q)$ is strictly proper. Either way, the transfer function on the right-hand side of \eqref{lefthandside} is stable and strictly proper and therefore it can be decomposed as
	\begin{equation}
		\label{tf}
		\frac{-\bar{\mathbf{B}}_{n+m}^{\textnormal{d}}(q)C_\textnormal{d}(q)}{\bar{A}_\textnormal{d}^{2}(q)\big(1+G_\textnormal{d}^*(q)C_\textnormal{d}(q)\big)} = \mathbf{g}_1 q^{-1}+\mathbf{g}_2 q^{-2}+\cdots,
	\end{equation}
	for some constant vectors $\mathbf{g}_1,\mathbf{g}_2,\dots$. Since $v(t_k)$ is white noise, the expectation of interest is given by
	\begin{equation}
		\mathbf{S}(-\hspace{-0.02cm}\bar{B},\hspace{-0.02cm}\bar{A})\mathbb{E}\Big\{\hspace{-0.04cm}\big[\hspace{-0.04cm}\hspace{-0.03cm}\left(\mathbf{g}_1 q^{\hspace{-0.03cm}-1}\hspace{-0.06cm}+\hspace{-0.03cm}\mathbf{g}_2 q^{-2}\hspace{-0.06cm}+\cdots\right)v(t_k)\big] v(t_k)\Big\} = \mathbf{0}, \notag
	\end{equation}
	which proves the result. \hspace*{\fill} \qed
\end{pf*}	

\bibliographystyle{plain}        
\bibliography{References}     

\begin{thebibliography}{10}

\bibitem{aastrom1970introduction}
K.~J. {\AA}str{\"o}m.
\newblock {\em Introduction to {S}tochastic {C}ontrol {T}heory}.
\newblock Academic Press, 1970.

\bibitem{dahleh2002lectures}
M.~Dahleh, M.~A. Dahleh, and G.~Verghese.
\newblock {\em Lectures on dynamic systems and control}.
\newblock Department of Electrical Engineering and Computer Science,
  Massachusetts Institute of Technology, 2002.

\bibitem{garnier2000bias}
H.~Garnier, M.~Gilson, and W.~X. Zheng.
\newblock A bias-eliminated least-squares method for continuous-time model
  identification of closed-loop systems.
\newblock {\em International Journal of Control}, 73(1):38--48, 2000.

\bibitem{garnier2008book}
H.~Garnier and L.~Wang~(Eds.).
\newblock {\em Identification of Continuous-time Models from Sampled Data}.
\newblock Springer, 2008.

\bibitem{garnier2014advantages}
H.~Garnier and P.~C. Young.
\newblock The advantages of directly identifying continuous-time transfer
  function models in practical applications.
\newblock {\em International Journal of Control}, 87(7):1319--1338, 2014.

\bibitem{gilson2003continuous}
M.~Gilson and H.~Garnier.
\newblock Continuous-time model identification of systems operating in
  closed-loop.
\newblock {\em IFAC Proceedings Volumes}, 36(16):405--410, 2003.

\bibitem{gilson2008instrumental}
M.~Gilson, H.~Garnier, P.~C. Young, and P.~M.~J. Van~den Hof.
\newblock Instrumental variable methods for closed-loop continuous-time model
  identification.
\newblock In {\em \textnormal{H. Garnier and L. Wang (Eds.). }Identification of
  {C}ontinuous-time {M}odels from {S}ampled {D}ata}, pages 133--160. Springer,
  2008.

\bibitem{gilson2011optimal}
M.~Gilson, H.~Garnier, P.~C. Young, and P.~M.~J. Van~den Hof.
\newblock Optimal instrumental variable method for closed-loop identification.
\newblock {\em IET Control Theory \& Applications}, 5(10):1147--1154, 2011.

\bibitem{gilson2005instrumental}
M.~Gilson and P.~M.~J. Van~den Hof.
\newblock Instrumental variable methods for closed-loop system identification.
\newblock {\em Automatica}, 41(2):241--249, 2005.

\bibitem{gonzalez2020consistent}
R.~A. Gonz{\'a}lez, C.~R. Rojas, S.~Pan, and J.~S. Welsh.
\newblock Consistent identification of continuous-time systems under multisine
  input signal excitation.
\newblock {\em Automatica}, 133, \textnormal{Article 109859}, 2020.

\bibitem{gonzalez2021srivc}
R.~A. Gonz{\'a}lez, C.~R. Rojas, S.~Pan, and J.~S. Welsh.
\newblock The {SRIVC} algorithm for continuous-time system identification with
  arbitrary input excitation in open and closed loop.
\newblock In {\em 60th IEEE Conference on Decision and Control (CDC)}, pages
  3004--3009, 2021.

\bibitem{gonzalez2021unstable}
R.~A. Gonz{\'a}lez, C.~R. Rojas, S.~Pan, and J.~S. Welsh.
\newblock Refined instrumental variable methods for unstable continuous-time
  systems in closed-loop.
\newblock {\em International Journal of Control}, pages 1--15, 2022.

\bibitem{goodwin2001control}
G.~C. Goodwin, S.~F. Graebe, and M.~E. Salgado.
\newblock {\em Control {S}ystem {D}esign}.
\newblock Prentice Hall, 2001.

\bibitem{goodwin2002bias}
G.~C. Goodwin and J.~S. Welsh.
\newblock Bias issues in closed loop identification with application to
  adaptive control.
\newblock {\em Communications in Information and Systems}, 2(4):349--370, 2002.

\bibitem{laurain2009refined}
V.~Laurain, M.~Gilson, and H.~Garnier.
\newblock Refined instrumental variable methods for identifying {H}ammerstein
  models operating in closed loop.
\newblock In {\em Proceedings of the 48h IEEE Conference on Decision and
  Control (CDC)}, pages 3614--3619, 2009.

\bibitem{lehmann1998theory}
E.~L. Lehmann and G.~Casella.
\newblock {\em Theory of {P}oint {E}stimation, \textnormal{2nd Edition}}.
\newblock Springer, 1998.

\bibitem{li2015closed}
Q.~Li, D.~Li, and L.~Cao.
\newblock Closed-loop identification of systems using hybrid {B}ox--{J}enkins
  structure and its application to {PID} tuning.
\newblock {\em Chinese {J}ournal of {C}hemical {E}ngineering},
  23(12):1997--2004, 2015.

\bibitem{ljung1998system}
L.~Ljung.
\newblock {\em System Identification: Theory for the User, \textnormal{2nd
  Edition}}.
\newblock Prentice-Hall, 1999.

\bibitem{mityagin2020zero}
B.~S. Mityagin.
\newblock The zero set of a real analytic function.
\newblock {\em Matematicheskie Zametki}, 107(3):473--475, 2020.

\bibitem{pan2020consistency}
S.~Pan, R.~A. Gonz{\'a}lez, J.~S. Welsh, and C.~R. Rojas.
\newblock Consistency analysis of the simplified refined instrumental variable
  method for continuous-time systems.
\newblock {\em Automatica}, 113, \textnormal{Article 108767}, 2020.

\bibitem{pan2020efficiency}
S.~Pan, J.~S. Welsh, R.~A. Gonz{\'a}lez, and C.~R. Rojas.
\newblock Efficiency analysis of the simplified refined instrumental variable
  method for continuous-time systems.
\newblock {\em Automatica}, 121, \textnormal{Article 109196}, 2020.

\bibitem{soderstrom1975ergodicity}
T.~S{\"o}derstr{\"o}m.
\newblock Ergodicity results for sample covariances.
\newblock {\em Problems of Control and Information Theory}, 4(2):131--138,
  1975.

\bibitem{soderstrom1983instrumental}
T.~S{\"o}derstr{\"o}m and P.~Stoica.
\newblock {\em Instrumental {V}ariable {M}ethods for {S}ystem
  {I}dentification}.
\newblock Springer, 1983.

\bibitem{soderstrom1984generic}
T.~S{\"o}derstr{\"o}m and P.~Stoica.
\newblock On the generic consistency of instrumental variable estimates.
\newblock {\em IFAC Proceedings Volumes}, 17(2):603--607, 1984.

\bibitem{soderstrom1988system}
T.~S{\"o}derstr{\"o}m and P.~Stoica.
\newblock {\em System {I}dentification}.
\newblock Prentice-Hall, 1989.

\bibitem{victor2017closed}
S.~Victor, A.~Diudichi, and P.~Melchior.
\newblock Closed-loop continuous-time model identification with noisy
  input-output.
\newblock {\em IFAC-PapersOnLine}, 50(1):12853--12858, 2017.

\bibitem{welsh2002finite}
J.~S. Welsh and G.~C. Goodwin.
\newblock Finite sample properties of indirect nonparametric closed-loop
  identification.
\newblock {\em IEEE Transactions on Automatic control}, 47(8):1277--1292, 2002.

\bibitem{young2008refined}
P.~C. Young.
\newblock The refined instrumental variable method.
\newblock {\em Journal Europ{\'e}en des Syst{\`e}mes Automatis{\'e}s},
  42(2-3):149--179, 2008.

\bibitem{young2009three}
P.~C. Young.
\newblock A three stage refined {IV} algorithm for closed loop identification
  and estimation.
\newblock Technical Report TR/210a, Faculty of Science and Technology,
  Lancaster University, 2009.

\bibitem{young2012recursive}
P.~C. Young.
\newblock {\em Recursive Estimation and Time-Series Analysis: An Introduction
  for the Student and Practitioner, \textnormal{2nd Edition}}.
\newblock Springer, 2012.

\bibitem{young2009simple}
P.~C. Young, H.~Garnier, and M.~Gilson.
\newblock Simple refined {IV} methods of closed-loop system identification.
\newblock In {\em 15th {IFAC} {S}ymposium on {S}ystem {I}dentification,
  \textnormal{Saint Malo, France}}, pages 1151--1156, 2009.

\bibitem{young1980refined}
P.~C. Young and A.~J. Jakeman.
\newblock Refined instrumental variable methods of recursive time-series
  analysis. {P}art {III}, {E}xtensions.
\newblock {\em International Journal of Control}, 31(4):741--764, 1980.

\end{thebibliography}
\end{document}